\batchmode
\makeatletter
\def\input@path{{"C:/Users/hsweatland/Documents/Papers/Journal Papers/[2024 CSL] DNN CBF/TAC Submission/"}}
\makeatother
\documentclass[english]{IEEEtran}
\pdfoutput=1
\usepackage[T1]{fontenc}
\usepackage[latin9]{inputenc}
\usepackage{color}
\usepackage{array}
\usepackage{multirow}
\usepackage{amsmath}
\usepackage{amsthm}
\usepackage{amssymb}
\usepackage{stackrel}
\usepackage{graphicx}
\usepackage{esint}

\makeatletter

\providecommand{\tabularnewline}{\\}

\theoremstyle{definition}
\newtheorem{defn}{\protect\definitionname}
\theoremstyle{definition}
\newtheorem{assumption}{Assumption}
\theoremstyle{remark}
\newtheorem{rem}{\protect\remarkname}
\theoremstyle{plain}
\newtheorem{lem}{\protect\lemmaname}
\theoremstyle{plain}
\newtheorem{thm}{\protect\theoremname}

\usepackage[noadjust]{cite}
\usepackage{multirow}

\makeatother

\usepackage{babel}
\providecommand{\definitionname}{Definition}
\providecommand{\lemmaname}{Lemma}
\providecommand{\remarkname}{Remark}
\providecommand{\theoremname}{Theorem}

\begin{document}
\global\long\def\R{\mathbb{R}}%
\global\long\def\r{\mathbb{R}}%
\global\long\def\S{{\cal S}}%

\title{Adaptive Deep Neural Network-Based Control Barrier Functions\thanks{H. M. Sweatland, O. S. Patil, and W. E. Dixon are with the Department
of Mechanical and Aerospace Engineering, University of Florida, Gainesville,
FL 32611, USA. Email: \{hsweatland, patilomkarsudhir, wdixon\}@ufl.edu.}\thanks{This research is supported in part by Office of Naval Research Grant
N00014-21-1-2481; AFOSR award number FA9550-19-1-0169. Any opinions,
findings and conclusions or recommendations expressed in this material
are those of the author(s) and do not necessarily reflect the views
of the sponsoring agency.}}
\author{Hannah M. Sweatland, Omkar Sudhir Patil, and Warren E. Dixon}
\maketitle
\begin{abstract}
Safety constraints of nonlinear control systems are commonly enforced
through the use of control barrier functions (CBFs). Uncertainties
in the dynamic model can disrupt forward invariance guarantees or
cause the state to be restricted to an overly conservative subset
of the safe set. In this paper, adaptive deep neural networks (DNNs)
are combined with CBFs to produce a family of controllers that ensure
safety while learning the system's dynamics in real-time without the
requirement for pre-training. By basing the least squares adaptation
law on a state derivative estimator-based identification error, the
DNN parameter estimation error is shown to be uniformly ultimately
bounded. The convergent bound on the parameter estimation error is
then used to formulate CBF-constraints in an optimization-based controller
to guarantee safety despite model uncertainty. Furthermore, the developed
method is extended for use under intermittent loss of state-feedback.
A switched systems analysis for CBFs is provided with a maximum dwell-time
condition during which the feedback can be unavailable. Comparative
simulation results demonstrate the ability of the developed method
to ensure safety in an adaptive cruise control problem and when feedback
is lost, unlike baseline methods. Results show improved performance
compared to baseline methods and demonstrate the ability of the developed
method to ensure safety in feedback-denied environments.
\end{abstract}

\section{Introduction\label{sec:Introduction}}

Control barrier functions (CBFs) enforce state constraints necessary
for safe operation of control systems \cite{Ames.Xu.ea2016,Ames.Coogan.ea2019},
but CBF-based control input constraints depend on the dynamic model
of the system. As a result, uncertainties in modeling the dynamics
can endanger safety. To address challenges posed by the modeling uncertainty,
robust safety methods can be used, where safety guarantees are provided
using the worst-case bounds on the uncertainty. However, robust methods
yield an overly conservative constraint on the control input that
restricts the state to a subset of the safe set. 

Adaptive CBFs have been developed to ensure the forward invariance
of a safe set through online parameter adaptation \cite{Taylor.Ames2020,Lopez.Slotine.ea2020,Cohen.Belta2022,Xiao.Belta.ea2022}.
Because the adaptive CBFs in both \cite{Taylor.Ames2020} and \cite{Lopez.Slotine.ea2020}
include the parameter estimation error, the state is restricted to
a subset of the safe set, dependent on the upper-bound of the estimation
error. Methods such as set membership identification \cite{Lopez.Slotine.ea2020},
integral concurrent learning \cite{Isaly.Patil.ea2021}, and parameter-adaptive
CBFs \cite{Xiao.Belta.ea2022} reduce the conservativeness with sufficient
data, but these methods involve a white-box approach where the uncertainty
is required to have a known structure based on traditional modeling
techniques. In contrast, recent results such as \cite{Srinivasan.Dabholkar.ea2020,Cheng.Orosz.ea2019,Isaly.Patil.ea2024}
use black-box models such as pre-trained deep neural networks (DNNs)
and Gaussian processes to identify uncertain dynamics using training
datasets and therefore reduce conservativeness; however, since these
methods are not adaptive, they result in static models that may become
obsolete over time. Moreover, the offline methods often require large
datasets that may not be available prior to execution in completely
unknown environments.

Other recent works reduce conservativeness of the CBF-based constraints
by combining CBFs with disturbance observers \cite{Agrawal.Panagou2022,Alan.Molnar.ea2023,Isaly.Patil.ea2024,Sun.Yang.ea2024}.
Disturbance observers are used to produce an estimate of the uncertainty
which is then used in the CBF constraint, expanding the state's operating
region when compared to a robust approach. Adaptive safety is achieved
in \cite{Isaly.Patil.ea2024} through a modular approach that can
combine a pre-trained DNN model with a robust integral sign of the
error (RISE)-based disturbance observer, eliminating conservativeness
of the safe set over time. Although disturbance observers can estimate
general nonlinear and time-varying uncertainties, the estimates are
only instantaneous and do not involve a model that can be used for
subsequent predictions. In contrast, models such as DNNs can extrapolate
through unexplored regions and can be employed to ensure safety under
intermittent loss of feedback. Therefore, instead of using disturbance
observers or pre-trained DNNs as in \cite{Agrawal.Panagou2022,Alan.Molnar.ea2023,Isaly.Patil.ea2024,Sun.Yang.ea2024}
it is desirable to construct adaptive CBFs using DNNs such as those
in \cite{Patil.Le.ea2022,Vacchini.Sacchi.ea2023,Li.Nguyen.ea2023,Patil.Griffis.ea2023,Griffis.Patil.ea.2023}
with analytic real-time adaptation laws without the need for pre-training.
Previous Lyapunov-based (Lb-) DNN adaptive controllers address the
trajectory tracking problem; however, the tracking error-based adaptation
laws in such results are not suitable for the adaptive safety problem
since safety does not typically require tracking error convergence.
Thus, in the developed work, a novel weight adaptation law is formulated
to instead yield function approximation error convergence.

In this paper, adaptive DNN CBFs (aDCBFs) are developed to ensure
safety while learning the system's uncertain dynamics in real-time.
This paper provides the first result combining CBFs with an adaptive
DNN that updates in real-time, eliminating the need for pre-training.
The DNN adaptation law is not based on the tracking error as in all
previous Lb-DNN literature. Instead, a least squares adaptation law
is designed by constructing an identification error. Since computing
an identification error requires state-derivative information, an
interlaced approach is used where a secondary state-derivative estimator
is combined with the adaptive DNN to generate the adaptation laws.
A combined Lyapunov-based analysis yields guarantees on the DNN function
approximation. The convergent upper-bound of the parameter estimation
error is then used to formulate candidate CBF-based constraints in
an optimization-based control law to guarantee the forward invariance
of the safe set, while reducing the conservative behavior often seen
in robust approaches. As a result, during intermittent loss of feedback,
the DNN can be used to make open-loop predictions that are used to
reformulate CBF-based constraints to guarantee safety. Thus, the developed
method can be used for safe operation of uncertain systems in environments
with feedback occlusion zones, where intermittent loss of feedback
typically occurs. A switched systems analysis for CBFs is provided
with a maximum dwell-time condition during which the feedback can
be unavailable. Comparative simulation results are presented to demonstrate
the performance of the developed method on two control systems with
baseline results in \cite{Ames.Xu.ea2016} and \cite{Isaly.Patil.ea2024}.

\section{Notation and Preliminaries\label{sec:Preliminaries}}

Let $\R_{\ge0}\triangleq\left[0,\infty\right)$, $\R_{>0}\triangleq\left(0,\infty\right)$,
and $\R^{n\times m}$ represent the space of $n\times m$ dimensional
matrices. The identity matrix of size $n$ is denoted by $I_{n}$.
The $p$-norm is denoted by $\left\Vert \cdot\right\Vert _{p}$, $\left\Vert \cdot\right\Vert $
is the 2-norm, and $\left\Vert \cdot\right\Vert _{F}$ is the Frobenius
norm defined as $\left\Vert \cdot\right\Vert _{F}\triangleq\left\Vert \text{vec}\left(\cdot\right)\right\Vert $,
where $\text{vec}\left(\cdot\right)$ denotes the vectorization operator.
Given some matrix $A\triangleq\left[a_{i,j}\right]\in\mathbb{R}^{n\times m}$,
where $a_{i,j}$ denotes the element in the $i^{th}$ row and $j^{th}$
column of $A$, the vectorization operator is defined as $\mathrm{vec}(A)\triangleq\left[a_{1,1},\ldots,a_{n,1},\ldots,a_{1,m},\ldots,a_{n,m}\right]{}^{\top}\in\mathbb{R}^{nm}$.
Let the notation $\left[d\right]$ be defined as $\left[d\right]\triangleq\left\{ 1,2,\ldots,d\right\} $
and, for vectors $x\in\mathbb{R}^{n}$ and $y\in\mathbb{R}^{m}$,
let $\left(x,y\right)\triangleq\left[x^{\top},y^{\top}\right]^{\top}$.
For a set $B\subset\R^{n}$, the boundary of $B$ is denoted $\partial B$,
the interior of $B$ is denoted $\text{int}\left(B\right)$, and an
open neighborhood about $B$ is denoted $\mathcal{N}\left(B\right)$.
A set-valued mapping $M:B\rightrightarrows\mathbb{R}^{m}$ associates
every point $x\in B$ with a set $M\left(x\right)\subset\mathbb{R}^{m}$. 

\subsection{Deep Neural Network Model\label{subsec:Deep-Neural-Network-1}}

For simplicity in the illustration, a fully-connected DNN will be
described here. The following control and adaptation law development
can be generalized for any network architecture $\Phi$ with a corresponding
Jacobian $\Phi^{\prime}$. The reader is referred to \cite{Patil.Le.ea.2022}
and \cite{Griffis.Patil.ea23_2} for extending the subsequent development
to ResNets and LSTMs, respectively. Let $\sigma\in\R^{L_{\textrm{in}}}$
denote the DNN input with size $L_{\textrm{in}}\in\mathbb{Z}_{>0}$,
and $\theta\in\R^{p}$ denote the vector of DNN parameters (i.e.,
weights and bias terms) with size $p\in\mathbb{Z}_{>0}$. A fully-connected
feedforward DNN $\Phi\left(\sigma,\theta\right)$ with output size
$L_{\textrm{out}}\in\mathbb{Z}_{>0}$ is defined using a recursive
relation $\Phi_{j}\in\R^{L_{j+1}}$ given by \cite{Patil.Le.ea2022}
\begin{eqnarray}
\Phi_{j} & \triangleq & \begin{cases}
V_{j}^{\top}\phi_{j}\left(\Phi_{j-1}\right), & j\in\left[k\right],\\
V_{j}^{\top}\sigma_{a}, & j=0,
\end{cases}\label{eq:phij_dnn}
\end{eqnarray}
where $\Phi\left(\sigma,\theta\right)=\Phi_{k}$ , and $\sigma_{a}\triangleq\left[\begin{array}{cc}
\sigma^{\top} & 1\end{array}\right]^{\top}$ denotes the augmented input that accounts for the bias terms, $k\in\mathbb{Z}_{>0}$
denotes the total number of hidden layers, $V_{j}\in\mathbb{R}^{L_{j}\times L_{j+1}}$
denotes the matrix of weights and biases, and $L_{j}\in\mathbb{Z}_{>0}$
denotes the number of nodes in the $j^{\textrm{th}}$ layer for all
$j\in\left\{ 0,\ldots,k\right\} $ with $L_{0}\triangleq L_{\textrm{in}}+1$
and $L_{k+1}=L_{\textrm{out}}$. The vector of smooth activation functions
is denoted by $\phi_{j}:\mathbb{R}^{L_{j}}\to\mathbb{R}^{L_{j}}$
for all $j\in\left[k\right]$. If the DNN involves multiple types
of activation functions at each layer, then $\phi_{j}$ may be represented
as $\phi_{j}\triangleq\left[\varsigma_{j,1}\begin{array}{ccc}
\ldots & \varsigma_{j,L_{j}-1} & \mathrm{1}\end{array}\right]^{\top}$, where $\varsigma_{j,i}:\mathbb{R}\to\mathbb{R}$ denotes the activation
function at the $i^{\mathrm{th}}$ node of the $j^{\mathrm{th}}$
layer. For the DNN architecture in (\ref{eq:phij_dnn}), the vector
of DNN weights is $\theta\triangleq\left[\begin{array}{ccc}
\mathrm{vec}(V_{0})^{\top} & \ldots & \mathrm{vec}(V_{k})^{\top}\end{array}\right]^{\top}$ with size $p=\Sigma_{j=0}^{k}L_{j}L_{j+1}$. The Jacobian of the
activation function vector at the $j^{\mathrm{th}}$ layer is denoted
by $\phi_{j}^{\prime}:\mathbb{R}^{L_{j}}\to\mathbb{R}^{L_{j}\times L_{j}}$,
and $\phi_{j}^{\prime}(y)\triangleq\left.\frac{\partial}{\partial z}\phi_{j}\left(z\right)\right|_{z=y}$,
$\forall y\in\mathbb{R}^{L_{j}}$. Let the Jacobian of the DNN with
respect to the weights be denoted by $\Phi^{\prime}\left(\sigma,\theta\right)\triangleq\frac{\partial}{\partial\theta}\Phi\left(\sigma,\theta\right)$,
which can be represented using $\Phi^{\prime}\left(\sigma,\theta\right)=\left[\begin{array}{cccc}
\Phi_{0}^{\prime}, & \Phi_{1}^{\prime}, & \ldots, & \Phi_{k}^{\prime}\end{array}\right]$, where $\Phi_{j}^{\prime}\triangleq\frac{\partial}{\partial\mathrm{vec}(V_{j})}\Phi\left(\sigma,\theta\right)$
for all $j\in\left[k\right]$. Then, using (\ref{eq:phij_dnn}) and
the property $\frac{\partial}{\partial\mathrm{vec}\left(B\right)}\mathrm{vec}\left(ABC\right)=C^{\top}\otimes A$
\cite[Proposition 7.1.9]{Bernstein2009} yields $\Phi_{0}^{\prime}=\left(\stackrel{\curvearrowleft}{\stackrel[l=1]{k}{\prod}}V_{l}^{\top}\phi_{l}^{\prime}\left(\Phi_{l-1}\right)\right)(I_{L_{1}}\otimes\sigma_{a}^{\top})$
and $\Phi_{j}^{\prime}=\left(\stackrel{\curvearrowleft}{\stackrel[l=j+1]{k}{\prod}}V_{l}^{\top}\phi_{l}^{\prime}\left(\Phi_{l-1}\right)\right)\left(I_{L_{j+1}}\otimes\phi_{j}^{\top}\left(\Phi_{j-1}\right)\right)$,
for all $j\in\left[k\right]$, where the notation $\stackrel{\curvearrowleft}{\prod}$
denotes the right-to-left matrix product operation, i.e., $\stackrel{\curvearrowleft}{\stackrel[p=1]{m}{\prod}}A_{p}=A_{m}\ldots A_{2}A_{1}$
and $\stackrel{\curvearrowleft}{\stackrel[p=a]{m}{\prod}}A_{p}=I$
if $a>m$, and $\otimes$ denotes the Kronecker product.

\section{Problem Formulation\label{sec:Problem-Formulation}}

\subsection{Dynamic Model\label{subsec:Dynamic-Model}}

Consider the nonlinear dynamic system 
\begin{align}
\dot{x} & =f\left(x\right)+g\left(x\right)u,\label{eq:system}
\end{align}
where $x\in\R^{n}$ denotes the state, $f:\R^{n}\rightarrow\R^{n}$
denotes an unknown continuously differentiable function, $u\in\Psi\subset\R^{m}$
denotes the control input, and $g:\R^{n}\rightarrow\R^{n\times m}$
denotes the known control effectiveness matrix, where $\Psi:\R^{n}\rightrightarrows\mathbb{R}^{m}$
denotes the set of admissible control inputs. The control objective
is to design a controller that ensures the forward invariance of a
safe set $\mathcal{S}\subset\R^{n}$ despite uncertainty in $f\left(x\right)$.
Forward invariance is a common safety objective because trajectories
beginning inside a forward invariant safe set will remain in the safe
set.

Given a controller $\kappa:\R^{n}\to\R^{m}$ with $\kappa\left(x\right)\in\Psi\left(x\right)$
for all $x\in\R^{n}$, we refer to the closed-loop dynamics defined
by (\ref{eq:system}) and $\kappa$ as $f_{cl}\left(x\right)\triangleq f\left(x\right)+g\left(x\right)\kappa\left(x\right)$.
A solution to the closed-loop dynamics $t\mapsto x\left(t\right)$
is complete if $\text{dom}x$ is unbounded and maximal if there is
no solution $y$ such that $x\left(t\right)=y\left(t\right)$ for
all $t\in\text{dom}x$, where $\text{dom}x$ is a proper subset of
$\text{dom}y$. The set $\mathcal{S}$ is forward pre-invariant for
the closed-loop dynamics $\dot{x}=f_{cl}\left(x\right)$ if, for each
$x_{0}\in\mathcal{S}$ and each maximal solution $x$ starting from
$x_{0}$, $x\left(t\right)\in\mathcal{S}$ for all $t\in\text{dom}x$
\cite[Definition 2.5]{Chai.Sanfelice2015}. The set is forward invariant
for the closed-loop dynamics if it is forward pre-invariant and additionally,
for each $x_{0}\in\mathcal{S}$, every maximal solution $x$ starting
from $x_{0}$ is complete \cite[Definition 2.6]{Chai.Sanfelice2015}.

\subsection{Control Barrier Functions (CBFs)\label{subsec:Control-Barrier-Functions}}

CBFs are a method used to encode a system's safety requirements. Using
the development in \cite{Isaly.Mamaghani.ea2024}, multiple scalar-valued
CBF candidates can be used to define the safe set.
\begin{defn}
\cite[Def. 1]{Isaly.Mamaghani.ea2024} A vector-valued function $B:\R^{n}\rightarrow\R^{d}$
is a CBF candidate defining a safe set $\mathcal{S}\subset\R^{n}$
if $\mathcal{S}=\left\{ x\in\R^{n}:B\left(x\right)\le0\right\} $,
where $B\left(x\right)\triangleq\left(B_{1}\left(x\right),B_{2}\left(x\right),\ldots,B_{d}\left(x\right)\right)$.
Also let $\mathcal{S}_{i}\triangleq\left\{ x\in\R^{n}:B_{i}\left(x\right)\le0\right\} $
and $M_{i}\triangleq\left\{ x\in\partial\mathcal{S}:B_{i}\left(x\right)=0\right\} $
for each $i\in\left[d\right]$.
\end{defn}
The state constraints defined by the CBF candidate can then be translated
to constraints on the control input, through the introduction of a
performance function $\gamma:\R^{n}\to\R^{d}$. The design parameter
$\gamma$ limits the worst-case growth of $B$ to ensure forward invariance
of $\mathcal{S}$ based on conditions derived in \cite{Maghenem.Sanfelice2021}.
\begin{defn}
\label{def:CBF}A continuously differentiable CBF candidate $B:\mathbb{R}^{n}\rightarrow\R^{d}$
defining the set $\mathcal{S}\subset\mathbb{R}^{n}$ is a CBF for
(\ref{eq:system}) and $\mathcal{S}$ on a set $\mathcal{O}\subset\R^{n}$
with respect to a function $\gamma:\R^{n}\to\R^{d}$ if 1) there exists
a neighborhood of the boundary of $\mathcal{S}$ such that $\mathcal{N}\left(\partial\mathcal{S}\right)\subset\mathcal{O}$,
2) the function $\gamma$ is such that, for each $i\in\left[d\right]$,
$\gamma_{i}\left(x\right)\ge0$ for all $x\in\mathcal{N}\left(M_{i}\right)\backslash\mathcal{S}_{i}$,
and 3) the set 
\begin{align}
K_{c}\left(x\right) & \triangleq\left\{ u\in\Psi\left(x\right):\nabla B^{\top}\left(x\right)\left(f\left(x\right)+g\left(x\right)u\right)\le-\gamma\left(x\right)\right\} \label{eq:Kc}
\end{align}
 is nonempty for every $x\in\mathcal{O}$.
\end{defn}
Since $f\left(x\right)$ is unknown, the inequality defining $K_{c}$
in (\ref{eq:Kc}) cannot be guaranteed to be satisfied without using
a conservative bound on the dynamics that would restrict the state's
operating region to a subset of the safe set. Thus, there is motivation
to develop an estimate of the uncertain dynamics to expand the operating
region. DNNs are a powerful tool that can be used to produce a real-time
approximation of $f\left(x\right)$.

\subsection{Deep Neural Network (DNN) Approximation\label{subsec:Deep-Neural-Network}}

Based on the universal function approximation theorem, DNNs can be
used to approximate continuous functions that lie on a compact set
\cite{Kidger.Lyons2020}. Lyapunov-based methods have been developed
to update the layer weights of a number of neural network (NN) architectures
including fully-connected DNNs \cite{Patil.Le.ea2022}, long short-term
memory NNs \cite{Griffis.Patil.ea23_2}, deep recurrent NNs \cite{Griffis.Patil.ea.2023},
and deep residual NNs (ResNets) \cite{Patil.Le.ea.2022}. 

On a compact set $\Omega\subset\R^{n}$, the uncertain dynamics in
(\ref{eq:system}) can be modeled as
\begin{align}
f\left(x\right) & =\Phi\left(x,\theta^{*}\right)+\varepsilon\left(x\right),\label{eq:DNN approximation}
\end{align}
where $\Phi:\R^{n}\times\R^{p}\rightarrow\R^{n}$ denotes the selected
DNN architecture, $\theta^{*}\in\R^{p}$ denotes a vector of ideal
weights defined as $\theta^{*}\triangleq\underset{\theta}{\arg\ \min}\ \underset{x\in\Omega}{\sup}\left(\left\Vert f\left(x\right)-\Phi\left(x,\theta\right)\right\Vert ^{2}+\sigma\left\Vert \theta\right\Vert ^{2}\right)$,
where $\sigma\in\R_{>0}$ is a regularizing constant, and $\varepsilon:\R^{n}\rightarrow\R^{n}$
denotes the unknown function reconstruction error. By the universal
function approximation property, for any prescribed $\overline{\varepsilon}\in\R_{>0}$,
there exist ideal DNN weights such that $\underset{x\in\Omega}{\sup}\left\Vert f\left(x\right)-\Phi\left(x,\theta^{*}\right)\right\Vert \le\overline{\varepsilon}$.
The function approximation error in (\ref{eq:DNN approximation})
satisfies $\underset{x\in\Omega}{\sup}\left\Vert \varepsilon\left(x\right)\right\Vert \le\overline{\varepsilon}$
on the compact domain $\Omega$. The subsequent CBF analysis ensures
the input to the DNN, $x$, remains in the forward invariant safe
set $\mathcal{S}\subseteq\Omega$ for all time, so the universal function
approximation property can be applied.

The DNN has a nested nonlinearly parameterized structure, so traditional
adaptive control techniques used for linearly parameterized systems
are not applicable. To help overcome the complexities introduced by
the nonlinearities, a first-order Taylor series approximation can
be used to estimate $\Phi\left(x,\theta^{*}\right)$ as \cite{Patil.Le.ea2022}
\begin{align}
\Phi\left(x,\theta^{*}\right) & =\Phi\left(x,\hat{\theta}\right)+\Phi^{\prime}\tilde{\theta}+\Delta_{O}^{2}\left(\left\Vert \tilde{\theta}\right\Vert \right),\label{eq:Taylor series}
\end{align}
where $\hat{\theta}\in\R^{p}$ denotes a vector composed of the adaptive
estimates the DNN layer weights that are generated using the subsequently
designed adaptation laws, $\Phi^{\prime}\in\R^{n\times p}$ denotes
the Jacobian of the DNN architecture defined as $\Phi^{\prime}\triangleq\frac{\partial\Phi\left(x,\hat{\theta}\right)}{\partial\hat{\theta}}$,
$\tilde{\theta}\in\R^{p}$ denotes the weight estimation error defined
as $\tilde{\theta}\triangleq\theta^{*}-\hat{\theta}$, and $\Delta_{O}^{2}:\R^{p}\rightarrow\R^{n}$
denotes higher-order terms. The following assumption is made to facilitate
the subsequent analysis.
\begin{assumption}
\label{asm:bounded ideal weights}There exists a known constant $\overline{\theta}\in\R_{>0}$
such that the unknown ideal weights can be bounded as $\left\Vert \theta^{*}\right\Vert \le\overline{\theta}$
\cite[Assumption 1]{Lewis.Yesildirek.ea1996}. Additionally, there
exists a known constant $\Xi\in\mathbb{R}_{>0}$ such that $\left\Vert f\left(x\right)-\Phi\left(x,\theta\right)\right\Vert ^{2}+\sigma\left\Vert \theta\right\Vert ^{2}$
is strictly convex with respect to $\theta$ for all $\theta\in\mathcal{B}\triangleq\left\{ \vartheta\in\mathbb{R}^{p}:\left\Vert \theta^{*}-\vartheta\right\Vert \leq\Xi\right\} $.
\end{assumption}
\begin{rem}
In Assumption \ref{asm:bounded ideal weights}, local convexity of
the regularized loss function $\left\Vert f\left(x\right)-\Phi\left(x,\theta\right)\right\Vert ^{2}+\sigma\left\Vert \theta\right\Vert ^{2}$
is assumed to facilitate convergence to a local minimum in the subsequent
analysis. A number of theoretical results in deep learning literature
that indicate that for some DNN architectures such as deep ResNets,
every local minimum is a global minimum \cite{Kawaguchi2016,Lu.Kawaguchi2017,Du.Lee.ea2018,Kawaguchi.Bengio2019}.
Furthermore, the regularizing term $\sigma\left\Vert \theta\right\Vert ^{2}$
assists in convexifying the loss function and mitigating practical
issues in deep learning such as overfitting \cite[Chapter 7]{Goodfellow2016}.
The strict convexity assumption on the loss function ensures that
$\theta^{*}$ is unique.
\end{rem}
Substituting the DNN estimate in (\ref{eq:DNN approximation}) and
Taylor series approximation in (\ref{eq:Taylor series}) into the
left-hand side of the inequality in (\ref{eq:Kc}) yields 
\begin{align}
\dot{B}\left(x,u\right) & =\nabla B^{\top}\left(x\right)\left(\Phi\left(x,\hat{\theta}\right)+\Phi^{\prime}\tilde{\theta}+\Delta+g\left(x\right)u\right),\label{eq:B_dot 2}
\end{align}
where $\Delta\in\R^{n}$ is defined as $\Delta\triangleq\Delta_{O}^{2}\left(\left\Vert \tilde{\theta}\right\Vert \right)+\varepsilon\left(x\right)$.
Although the DNN approximation alone is less conservative than bounding
the entire uncertainty, (\ref{eq:B_dot 2}) is still composed of the
unknown terms $\tilde{\theta}$ and $\Delta$. Assumption \ref{asm:bounded ideal weights}
could be used to bound $\tilde{\theta}$ in the CBF constraint in
(\ref{eq:B_dot 2}), but instead we introduce an adaptive identifier
in the following subsection to further reduce conservative behavior
due to the uncertainty in $\tilde{\theta}$.

\subsection{Adaptive DNN-Based Identifier Design\label{subsec:Observer-Design}}

All previous Lb-DNN-based adaptive control results update the DNN
weights using the tracking error \cite{Patil.Le.ea2022,Griffis.Patil.ea23_2,Patil.Griffis.ea2023,Hart.patil.ea2023,Griffis.Patil.ea.2023,Patil.Le.ea.2022};
however, the objective in those results is to track a desired trajectory.
To achieve adaptive safety, the adaptive weight updates need to be
performed with system identification as the objective. Therefore,
a least squares weight adaptation law is introduced to adaptively
identify the system dynamics based on an identification error. Performing
least squares-based real-time identification is challenging for continuous-time
systems because it requires state-derivative information which is
often unknown or noisy. Therefore, we introduce a high-gain state-derivative
estimator defined as
\begin{align}
\dot{\hat{x}} & =\hat{f}+g\left(x\right)u+k_{x}\tilde{x},\label{eq:x hat dot}
\end{align}
\begin{align}
\dot{\hat{f}} & =k_{f}\left(\dot{\tilde{x}}+k_{x}\tilde{x}\right)+\tilde{x},\label{eq:fhat dot}
\end{align}
where $\hat{x},\hat{f}\in\R^{n}$ denote the observer estimates of
$x$ and $f$, respectively, $k_{x},k_{f}\in\R_{>0}$ are positive
constant observer gains, and observer errors $\tilde{x},\tilde{f}\in\mathbb{\R}^{n}$
are defined as $\tilde{x}\triangleq x-\hat{x}$ and $\tilde{f}\triangleq f\left(x\right)-\hat{f}$,
respectively. Since state feedback is available, $\tilde{x}$ is known,
but $\dot{\tilde{x}}$ is unknown. An implementable form of $\hat{f}$
can be found by integrating both sides of $\dot{\hat{f}}$ in (\ref{eq:fhat dot})
to yield $\hat{f}\left(t\right)=\hat{f}\left(t_{0}\right)+k_{f}\tilde{x}\left(t\right)-k_{f}\tilde{x}\left(t_{0}\right)+\intop_{t_{0}}^{t}\left(k_{f}k_{x}+1\right)\tilde{x}\left(\tau\right)d\tau$.
Taking the time derivative of the definitions of $\tilde{x}$ and
$\tilde{f}$ and substituting (\ref{eq:x hat dot}) and (\ref{eq:fhat dot})
yields 
\begin{align}
\dot{\tilde{x}} & =\tilde{f}-k_{x}\tilde{x},\label{eq:x tilde dot}
\end{align}
\begin{align}
\dot{\tilde{f}} & =\dot{f}-k_{f}\tilde{f}-\tilde{x},\label{eq:f tilde dot}
\end{align}
where $\dot{f}\left(x\right)\triangleq\nabla f^{\top}\left(x\right)\dot{x}$.
The following lemma is provided to establish the boundedness of $\dot{f}$
based on the continuous differentiability of $f$ under common assumptions
in CBF literature. 
\begin{lem}
\label{lem:boundedness of f dot}Consider the function $f$, a continuous
controller $\kappa\in\Psi$ and the set $\mathcal{S}\subset\mathbb{R}^{n}$.
Based on the continuous differentiability of $f$, the continuity
$\kappa$, and the boundedness of $x$ on $\mathcal{S}$, the signals
$f$ and $\kappa$ are bounded on $\mathcal{S}$. Therefore, there
exists a known constant $\bar{\dot{f}}\in\R_{>0}$ such that $\left\Vert \dot{f}\left(x\right)\right\Vert \le\bar{\dot{f}}$
for all $x\in\mathcal{S}$.
\end{lem}
\begin{IEEEproof}
The safe set $\mathcal{S}$ is compact because it is a closed subset
of the compact set $\Omega$. Because of the continuity of $f$ and
the fact that $\mathcal{S}$ is compact, there exists a known constant
$\overline{f}\in\R{}_{\ge0}$ such that $\left\Vert f\left(x\right)\right\Vert \le\overline{f}$
for all $x\in\mathcal{S}$. The controller $\kappa$ and control effectiveness
$g$ are continuous and therefore bounded for all $x\in\mathcal{S}$.
Thus, because $\dot{x}=f\left(x\right)+g\left(x\right)\kappa\left(x\right)$,
it follows that $\dot{x}$ is bounded for all $x\in\mathcal{S}$ based
on the boundedness of $f$, $x$, and $\kappa$ on $\mathcal{S}$.
The function $f$ is continuously differentiable, so $\nabla f$ is
bounded on $\mathcal{S}$. Since $\dot{f}\left(x\right)=\nabla f^{\top}\left(x\right)\dot{x}$,
there exists a constant $\bar{\dot{f}}\in\R_{>0}$ such that $\left\Vert \dot{f}\left(x\left(t\right)\right)\right\Vert \le\bar{\dot{f}}$
for all $x\in\mathcal{S}$.
\end{IEEEproof}
\begin{figure}
\centering{}\includegraphics[width=1\columnwidth]{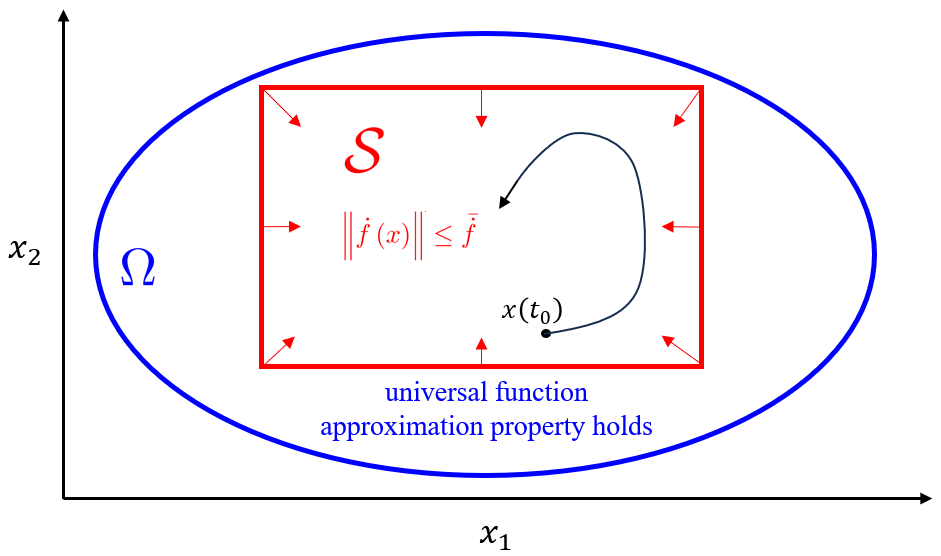}\caption{\label{fig:sets}An illustration of the sets $\Omega$ and ${\cal S}$
in $\mathbb{R}^{2}$. On the blue set $\Omega$ the universal function
approximation property holds. Flows generated by the CBF are constrained
to the red set ${\cal S}$, where $\left\Vert \dot{f}\left(x\right)\right\Vert \le\bar{\dot{f}}$.}
\end{figure}
Figure \ref{fig:sets} provides a visualization of the sets $\Omega$
and ${\cal S}$. The CBF constraints restrict the state trajectories
to the safe set ${\cal S}$, where $\left\Vert \dot{f}\left(x\right)\right\Vert \le\bar{\dot{f}}$.
Because ${\cal S}\subseteq\Omega$, the universal function approximation
property of DNNs holds on ${\cal S}$.

Based on the subsequent analysis, the DNN adaptation law $\dot{\hat{\theta}}\in\R^{p}$
is designed as 
\begin{align}
\dot{\hat{\theta}} & =\text{proj}\left(\Gamma\left(-k_{\theta}\hat{\theta}+\alpha\Phi^{\prime\top}\left(x,\hat{\theta}\right)\left(\hat{f}-\Phi\left(x,\hat{\theta}\right)\right)\right)\right),\label{eq:theta hat dot}
\end{align}
where\textcolor{blue}{{} }$k_{\theta},\alpha\in\R_{>0}$ denote constant
gains and the projection operator $\text{proj}\left(\cdot\right)$
is defined as in \cite[Appendix E]{Krstic1995} and ensures that $\hat{\theta}\left(t\right)\in{\cal B}$.
The term $\Gamma\in\R^{p\times p}$ denotes a symmetric positive-definite
time-varying least squares adaptation gain matrix that is a solution
to 
\begin{align}
\frac{d}{dt}\Gamma^{-1} & =-\beta\left(t\right)\Gamma^{-1}+\Phi^{\prime\top}\left(x,\hat{\theta}\right)\Phi^{\prime}\left(x,\hat{\theta}\right),\label{eq:derivative Gamma}
\end{align}
with the bounded-gain time-varying forgetting factor $\beta:\R_{\ge0}\rightarrow\R_{\ge0}$
designed as $\beta\left(t\right)\triangleq\beta_{0}\left(1-\frac{\left\Vert \Gamma\right\Vert }{\kappa_{0}}\right),$
where $\beta_{0},\kappa_{0}\in\r_{>0}$ are user-defined constants
that denote the maximum forgetting rate and the bound prescribed on
$\left\Vert \Gamma\right\Vert $, respectively. The adaptation gain
matrix is initialized to be positive-definite such that $\left\Vert \Gamma\left(t_{0}\right)\right\Vert <\kappa_{0}$,
and $\Gamma\left(t\right)$ remains positive-definite for all $t\in\R_{\ge0}$
\cite{Slotine1989}. Because $\Gamma\left(t\right)$ is positive-definite,
there exists a constant $\kappa_{1}\in\mathbb{R}_{>0}$ such that
$\lambda_{\text{min}}\left(\Gamma\left(t\right)\right)\ge\kappa_{1}$
for all $t\in\mathbb{R}_{\geq t_{0}}$. If $\Phi^{\prime}\left(x,\hat{\theta}\right)$
satisfies the persistence of excitation (PE) condition, meaning there
exist constants $\varphi_{1},\varphi_{2}\in\R_{>0}$ such that $\varphi_{1}I_{p}\le\int_{t_{1}}^{t_{1}+T}\Phi^{\prime\top}\left(x\left(\tau\right),\hat{\theta}\left(\tau\right)\right)\Phi^{\prime}\left(x\left(\tau\right),\hat{\theta}\left(\tau\right)\right)d\tau\le\varphi_{2}I_{p}$
for all $t_{1}\in\R_{\ge0}$ and $T\in\R_{>0}$, it can be shown that
$\beta_{1}>0$ \cite[Sec. 4.2]{Slotine1989}, where $\beta_{1}\in\R_{\ge0}$
is a constant such that $\beta\ge\beta_{1}$.

\section{Stability Analysis\label{sec:Stability-Analysis}}

Taking the time derivative of $\tilde{\theta}$, adding and subtracting
$f$, and substituting (\ref{eq:DNN approximation}) and (\ref{eq:Taylor series})
into (\ref{eq:theta hat dot}), the parameter estimation error dynamics
are given as 
\begin{align}
\dot{\tilde{\theta}} & =-\text{proj}\left(\Gamma\left(k_{\theta}\tilde{\theta}+\alpha\Phi^{\prime\top}\left(x,\hat{\theta}\right)\left(\Phi^{\prime}\tilde{\theta}+\Delta-\tilde{f}\right)-k_{\theta}\theta^{*}\right)\right).\label{eq:weight update law}
\end{align}
The subsequent Lyapunov-based stability analysis demonstrates the
convergence properties of (\ref{eq:x tilde dot}), (\ref{eq:f tilde dot}),
and (\ref{eq:weight update law}).

To facilitate the stability analysis, let $z\triangleq\left[\tilde{x}^{\top}\ \tilde{f}^{\top}\ \tilde{\theta}^{\top}\right]^{\top}\in\R^{2n+p}$
denote the concatenated state vector. Let the Lyapunov function candidate
$V:\R^{2n+p}\rightarrow\R$ be defined as
\begin{align}
V\left(z\right) & \triangleq\frac{1}{2}\tilde{x}^{\top}\tilde{x}+\frac{1}{2}\tilde{f}^{\top}\tilde{f}+\frac{1}{2}\tilde{\theta}^{\top}\Gamma^{-1}\tilde{\theta},\label{eq:Lyapunov function}
\end{align}
which can be bounded as 
\begin{align}
\lambda_{1}\left\Vert z\right\Vert ^{2} & \le V\left(z\right)\le\lambda_{2}\left\Vert z\right\Vert ^{2},\label{eq:Lyapunov inequality}
\end{align}
where $\lambda_{1}\triangleq\min\left\{ \frac{1}{2},\frac{1}{2\kappa_{0}}\right\} $,
$\lambda_{2}\triangleq\max\left\{ \frac{1}{2},\frac{1}{2\kappa_{1}}\right\} $.
Taking the time-derivative of (\ref{eq:Lyapunov function}), substituting
(\ref{eq:x tilde dot}), (\ref{eq:f tilde dot}), (\ref{eq:derivative Gamma}),
and (\ref{eq:weight update law}), and applying the property of projection
operators $-\tilde{\theta}^{\top}\Gamma^{-1}\text{proj}\left(\mu\right)\le-\tilde{\theta}^{\top}\Gamma^{-1}\mu$
\cite[Lemma E.1.IV]{Krstic1995}, yields
\begin{align}
\dot{V} & \le-k_{x}\left\Vert \tilde{x}\right\Vert ^{2}-k_{f}\left\Vert \tilde{f}\right\Vert ^{2}+\tilde{f}^{\top}\dot{f}\nonumber \\
 & -\left(\frac{\beta_{1}}{2\kappa_{0}}+k_{\theta}\right)\left\Vert \tilde{\theta}\right\Vert ^{2}-\left(\alpha-\frac{1}{2}\right)\tilde{\theta}^{\top}\Phi^{\prime\top}\left(x,\hat{\theta}\right)\Phi^{\prime}\left(x,\hat{\theta}\right)\tilde{\theta}\nonumber \\
 & +\alpha\tilde{\theta}^{\top}\Phi^{\prime\top}\left(x,\hat{\theta}\right)\left(\tilde{f}-\Delta\right)+k_{\theta}\tilde{\theta}^{\top}\theta^{*}.\label{eq:V dot}
\end{align}
Because $f$ and $\Phi$ are continuously differentiable $\left\Vert \Delta\right\Vert \le c_{1}$
and $\left\Vert \Phi^{\prime}\left(x,\hat{\theta}\right)\right\Vert _{F}\le c_{2}$
when $z\in\mathcal{D}\triangleq\left\{ \zeta\in\R^{2n+p}:\left\Vert \zeta\right\Vert \le\chi\right\} $,
where $c_{1},c_{2},\chi\in\R_{>0}$ are known constants. Recall that
by Lemma \ref{lem:boundedness of f dot}, there exists a known bound
$\overline{\dot{f}}\in\R_{>0}$ such that $\left\Vert \dot{f}\right\Vert \le\overline{\dot{f}}$
for all $x\in\mathcal{S}$. Using Young's Inequality and Assumption
\ref{asm:bounded ideal weights}, $\tilde{\theta}^{\top}\Phi^{\prime\top}\left(x,\hat{\theta}\right)\left(\tilde{f}-\Delta\right)\le c_{2}\left\Vert \tilde{\theta}\right\Vert ^{2}+\frac{c_{2}}{2}\left\Vert \tilde{f}\right\Vert ^{2}+\frac{c_{2}c_{1}^{2}}{2}$,
$\tilde{f}^{\top}\dot{f}\le\frac{\overline{\dot{f}}}{2}\left\Vert \tilde{f}\right\Vert ^{2}+\frac{\overline{\dot{f}}}{2}$,
and $k_{\theta}\tilde{\theta}^{\top}\theta^{*}\le\frac{k_{\theta}}{2}\left\Vert \tilde{\theta}\right\Vert ^{2}+\frac{k_{\theta}}{2}\overline{\theta}^{2}$,
so (\ref{eq:V dot}) can be further bounded as
\begin{align}
\dot{V} & \le-\lambda_{3}\left\Vert z\right\Vert ^{2}+C-\left(\alpha-\frac{1}{2}\right)\tilde{\theta}^{\top}\Phi^{\prime\top}\left(x,\hat{\theta}\right)\Phi^{\prime}\left(x,\hat{\theta}\right)\tilde{\theta},\label{eq:V dot 1}
\end{align}
where $\lambda_{3}\triangleq\min\left\{ k_{x},\ k_{f}-\frac{\overline{\dot{f}}}{2}-\frac{c_{2}}{2},\ \frac{\beta_{1}}{2\kappa_{0}}+\frac{k_{\theta}}{2}-c_{2}\right\} $
and $C\triangleq\frac{\overline{\dot{f}}+c_{2}c_{1}^{2}+k_{\theta}\overline{\theta}^{2}}{2}$.
Additionally, let $\mathcal{Q}\triangleq\left\{ \zeta\in\R^{2n+p}:\left\Vert \zeta\right\Vert \le\sqrt{\frac{\lambda_{1}}{\lambda_{2}}\chi^{2}-\frac{C}{\lambda_{3}}}\right\} $,
which is defined to initialize $z$ in the subsequent analysis. The
following theorem provides conditions under which the adaptation law
in (\ref{eq:theta hat dot}) yields parameter estimation error convergence.
\begin{thm}
\label{thm: theta tilde convergence}Let $t\mapsto x\left(t\right)$
be such that $x\left(t_{0}\right)\in\text{int}\left(\S\right)$\footnote{The initial conditions $x\left(t_{0}\right)$ is considered to be
in the interior of ${\cal S}$ to ensure ${\cal I}$ is not measure-zero,
thus ruling out solutions that instantly escape ${\cal S}$.} and there exists a time interval $\mathcal{I}\triangleq\left[t_{0},t_{\mathcal{I}}\right)$
such that $x\left(t\right)\in\mathcal{S}$ for all $t\in\mathcal{I}$.
If Assumption \ref{asm:bounded ideal weights} is satisfied, $\kappa$
is continuous, and $\chi>\sqrt{\frac{\lambda_{2}C}{\lambda_{1}\lambda_{3}}}$,
then the weight update law in (\ref{eq:theta hat dot}) ensures that
$\left\Vert \tilde{\theta}\left(t\right)\right\Vert \le\tilde{\theta}_{UB}\left(t\right)$
for all $t\in\mathcal{I}$, where 
\begin{align*}
\tilde{\theta}_{UB}\left(t\right) & \triangleq\sqrt{\frac{\lambda_{2}}{\lambda_{1}}\left\Vert z\left(t_{0}\right)\right\Vert ^{2}e^{-\frac{\lambda_{3}}{\lambda_{2}}t}+\frac{\lambda_{2}C}{\lambda_{1}\lambda_{3}}\left(1-e^{-\frac{\lambda_{3}}{\lambda_{2}}t}\right)},
\end{align*}
 provided $z\left(t_{0}\right)\in\mathcal{Q}$, $\hat{\theta}\left(t_{0}\right)\in{\cal B}$,
$\lambda_{3}>0$, and $\alpha>\frac{1}{2}$.
\end{thm}
\begin{IEEEproof}
From the Lyapunov function candidate in (\ref{eq:Lyapunov function})
and the inequalities in (\ref{eq:Lyapunov inequality}) and (\ref{eq:V dot 1}),
$\dot{V}$ can be further bounded as $\dot{V}\le-\frac{\lambda_{3}}{\lambda_{2}}V+C$,
for all $z\in{\cal D}$ and $t\in{\cal I}$ if the gain conditions
are satisfied. Solving the differential inequality over the time interval
${\cal I}$ yields 
\begin{align}
V\left(z\left(t\right)\right) & \le V\left(z\left(t_{0}\right)\right)e^{-\frac{\lambda_{3}}{\lambda_{2}}t}+\frac{\lambda_{2}C}{\lambda_{3}}\left(1-e^{-\frac{\lambda_{3}}{\lambda_{2}}t}\right),\label{eq:UUB bound}
\end{align}
for all $z\in{\cal D}$. From (\ref{eq:Lyapunov function}) and (\ref{eq:UUB bound}),
it follows that 
\begin{align}
\left\Vert z\left(t\right)\right\Vert  & \le\sqrt{\frac{\lambda_{2}}{\lambda_{1}}\left\Vert z\left(t_{0}\right)\right\Vert ^{2}e^{-\frac{\lambda_{3}}{\lambda_{2}}t}+\frac{\lambda_{2}C}{\lambda_{1}\lambda_{3}}\left(1-e^{-\frac{\lambda_{3}}{\lambda_{2}}t}\right)},\label{eq:z bound}
\end{align}
for all $z\in{\cal D}$ and $t\in{\cal I}$. To ensure $z\left(t\right)\in{\cal D}$
for all $t\in{\cal I}$, further upper-bounding (\ref{eq:z bound})
yields $\left\Vert z\left(t\right)\right\Vert \le\sqrt{\frac{\lambda_{2}}{\lambda_{1}}\left\Vert z\left(t_{0}\right)\right\Vert ^{2}+\frac{\lambda_{2}C}{\lambda_{1}\lambda_{3}}}$
for all $t\in{\cal I}$; hence, $z\left(t\right)\in{\cal D}$ always
holds if $\sqrt{\frac{\lambda_{2}}{\lambda_{1}}\left\Vert z\left(t_{0}\right)\right\Vert ^{2}+\frac{\lambda_{2}C}{\lambda_{1}\lambda_{3}}}\le\chi$,
which is guaranteed if $\left\Vert z\left(t_{0}\right)\right\Vert \le\sqrt{\frac{\lambda_{1}}{\lambda_{2}}\chi^{2}-\frac{C}{\lambda_{3}}}$,
i.e., $z\left(t_{0}\right)\in{\cal Q}$. Thus, trajectories of $z$
do not escape ${\cal D}$ if $z$ is initialized in ${\cal Q}$. Additionally,
because $\left\Vert \tilde{\theta}\right\Vert \le\left\Vert z\right\Vert $,
(\ref{eq:z bound}) implies 
\begin{align}
\left\Vert \tilde{\theta}\left(t\right)\right\Vert  & \le\sqrt{\frac{\lambda_{2}}{\lambda_{1}}\left\Vert z\left(t_{0}\right)\right\Vert ^{2}e^{-\frac{\lambda_{3}}{\lambda_{2}}t}+\frac{\lambda_{2}C}{\lambda_{1}\lambda_{3}}\left(1-e^{-\frac{\lambda_{3}}{\lambda_{2}}t}\right)},\label{eq:theta bound}
\end{align}
for all $t\in\mathcal{I}$ and $z\in{\cal D}$ if $z\left(t_{0}\right)\in\mathcal{Q}$.
For the initial conditions to be feasible, ${\cal Q}$ is required
to be non-empty, which is ensured by selecting $\chi>\sqrt{\frac{\lambda_{2}C}{\lambda_{1}\lambda_{3}}}$.
\end{IEEEproof}
The bound in (\ref{eq:theta bound}) cannot be implemented without
information about the concatenated initial state $z\left(t_{0}\right)$.
Since the state information is available, $\hat{x}$ is initialized
such that $\tilde{x}\left(t_{0}\right)=0$. While $\tilde{f}$ and
$\tilde{\theta}$ are unknown, each have known bounds. Recall that
because of the continuity of $f$ and the fact that $\S$ is a closed
subset of the compact set $\Omega$ and is therefore compact, there
exists a known constant $\overline{f}\in\R{}_{\ge0}$ such that $\left\Vert f\left(x\right)\right\Vert \le\overline{f}$
for all $x\in{\cal S}$. Thus, it follows that $\tilde{f}$ is bounded.
The bound on $\tilde{\theta}$ is a result of the projection operator
in (\ref{eq:weight update law}). If $\left\Vert \hat{f}\left(t_{0}\right)\right\Vert \le\overline{f}$
and $\hat{\theta}\left(t_{0}\right)\in{\cal B}$, then $\left\Vert \tilde{f}\left(t_{0}\right)\right\Vert \le2\overline{f}$
and $\left\Vert \tilde{\theta}\left(t_{0}\right)\right\Vert \le\Xi$.
Therefore, there exists a known constant ${\cal Z}\in\R_{>0}$ such
that $\left\Vert z\left(t_{0}\right)\right\Vert \le{\cal Z}\triangleq\sqrt{\Xi^{2}+4\overline{f}^{2}}$
and (\ref{eq:theta bound}) can be further bounded as
\begin{align}
\left\Vert \tilde{\theta}\left(t\right)\right\Vert  & \le\sqrt{\frac{\lambda_{2}}{\lambda_{1}}{\cal Z}^{2}e^{-\frac{\lambda_{3}}{\lambda_{2}}t}+\frac{\lambda_{2}C}{\lambda_{1}\lambda_{3}}\left(1-e^{-\frac{\lambda_{3}}{\lambda_{2}}t}\right)},\label{eq:theta bound-1}
\end{align}
for all $t\in\mathcal{I}$ if $z\left(t_{0}\right)\in\mathcal{Q}$.

Because the bound in (\ref{eq:theta bound-1}) may initially be more
conservative than $\Xi$, we design a function $\chi_{\theta}\in\r_{>0}$
as 
\begin{align}
\chi_{\theta} & \triangleq\min\left\{ \Xi,\sqrt{\frac{\lambda_{2}}{\lambda_{1}}{\cal Z}^{2}e^{-\frac{\lambda_{3}}{\lambda_{2}}t}+\frac{\lambda_{2}C}{\lambda_{1}\lambda_{3}}\left(1-e^{-\frac{\lambda_{3}}{\lambda_{2}}t}\right)}\right\} ,\label{eq:chi}
\end{align}
such that $\left\Vert \tilde{\theta}\left(t\right)\right\Vert \le\chi_{\theta}$
for all time. When the observer gains $k_{x}$ and $k_{f}$ are selected
to be sufficiently high, $\lambda_{3}=\frac{\beta_{1}}{2\kappa_{0}}+\frac{k_{\theta}}{2}-c_{2}$,
which implies the rate of convergence in (\ref{eq:theta bound-1})
depends primarily on $\beta_{1}$ and $k_{\theta}$. Thus, when the
PE condition is satisfied, $\beta_{1}>0$, resulting in a larger $\lambda_{3}$
which implies $\chi_{\theta}$ converges faster and to a smaller value.
When the PE condition is not satisfied, the gain $k_{\theta}$ helps
achieve the uniform ultimate boundedness of $\tilde{\theta}$ based
on sigma modification; however, selection of a larger $k_{\theta}$
yields a larger $C$, worsening parameter estimation performance.
By substituting the developed upper-bound of the parameter estimation
error in (\ref{eq:chi}) into (\ref{eq:B_dot 2}) and recalling $\left\Vert \Delta\right\Vert \le c_{1}$,
a new CBF notion composed of only the known signals can be defined. 

\begin{defn}
\label{def:aDNNCBF}A continuously differentiable CBF candidate $B:\R^{n}\rightarrow\R^{d}$
defining the set ${\cal S}\subseteq\Omega$ is an \textit{adaptive
DNN CBF (aDCBF) }for the dynamics in (\ref{eq:system}) and safe set
${\cal S}$ on a set ${\cal O}\subset\R^{n}$ with respect to $\gamma:\R^{n}\rightarrow\R^{d}$
if 1) there exists a neighborhood of the boundary of ${\cal S}$ such
that ${\cal N}\left(\partial{\cal S}\right)\subset{\cal O}$, 2) for
each $i\in\left[d\right]$, $\gamma_{i}\left(x\right)\ge0$ for all
$x\in{\cal N}\left(M_{i}\right)\backslash{\cal S}_{i}$, and 3) the
set 
\begin{align*}
K_{d}\left(x\right) & \triangleq\biggl\{ u\in\Psi:\left\Vert \nabla B^{\top}\left(x\right)\Phi^{\prime}\right\Vert \left(\chi_{\theta}+c_{1}\right)\\
 & +\nabla B^{\top}\left(x\right)\left(\Phi\left(x,\hat{\theta}\right)+g\left(x\right)u\right)\le-\gamma\left(x\right)\biggr\},
\end{align*}
is nonempty for all $x\in{\cal O}$.
\end{defn}
The set $K_{d}$ represents the set of control inputs that will render
the set ${\cal S}$ forward invariant. A selection of $K_{d}$ that
minimizes some cost function can be made at each $x\in{\cal O}$ using
an optimization-based control law. 

The controller $\kappa^{*}:\R^{n}\rightarrow\Psi$ is defined as
\begin{align}
\kappa^{*}\left(x\right) & \triangleq\underset{u\in\Psi}{\arg\min}\ Q\left(x,u\right),\label{eq:kappa star}\\
 & \ \text{s.t.}\ {\cal C}_{F}\left(x,u\right)\le0,\nonumber 
\end{align}
where $Q\left(x,u\right):\R^{n}\times\Psi\rightarrow\R$ is a cost
function typically selected as $\left\Vert u-u_{nom}\right\Vert ^{2}$,
$u_{nom}\in\Psi$ is a nominal continuous control input, and ${\cal C}_{F}:\R^{n}\times\Psi\rightarrow\R^{d}$
are the constraints on the control input. By choosing ${\cal C}_{F}\left(x,u\right)\triangleq\left\Vert \nabla B^{\top}\left(x\right)\Phi^{\prime}\right\Vert \left(\chi_{\theta}+c_{1}\right)+\nabla B^{\top}\left(x\right)\left(\Phi\left(x,\hat{\theta}\right)+g\left(x\right)u\right)+\gamma\left(x\right)$,
the optimization problem yields a controller $\kappa^{*}\left(x\right)\in K_{d}$
for all $x\in{\cal O}$ assuming $K_{d}$ is nonempty on the set ${\cal O}$.
To obtain an implementable form of the controller, we impose the following
conditions on the set of admissible control inputs $\Psi.$
\begin{assumption}
\label{asm:psi}There exists a function $\psi:\mathbb{R}^{n}\times\mathbb{R}^{m}\rightarrow\mathbb{R}^{s}$
such that $\Psi\left(x\right)=\left\{ u\in\mathbb{R}^{m}:\psi\left(x,u\right)\le0\right\} $
for all $x\in\mathbb{R}^{n}$. Additionally, for each $r\in\left[s\right]$,
the function $u\mapsto\psi_{r}\left(x,u\right)$ is convex on $K_{d}$
and $\left(x,u\right)\mapsto\psi\left(x,u\right)$ is continuous on
${\cal O}\times\mathbb{R}^{m}$.
\end{assumption}
The following lemma of \cite[Theorem 4]{Isaly.Mamaghani.ea2024} provides
conditions that result in a continuous $\kappa^{*}$. 

\begin{lem}
\label{lem:continuity of k*}\cite[Theorem 4]{Isaly.Mamaghani.ea2024}
Let ${\cal C}:\mathbb{R}^{n}\times\R^{m}\to\R^{h}$ be continuous
on ${\cal O}\times\mathbb{R}^{m}$, and, for each $g\in\left[h\right]$,
let $u\mapsto{\cal C}_{g}\left(x,u\right)$ be convex on the set $K\left(x\right)\triangleq\left\{ u\in\mathbb{R}^{m}:{\cal C}_{g}\left(x,u\right)\le0,\forall g\in\left[h\right]\right\} $.
Suppose $Q:\mathbb{R}^{n}\times\R^{m}\to\R$ is continuous and for
each $x\in{\cal O}$, $u\mapsto Q\left(x,u\right)$ is strictly convex
and inf-compact\footnote{A function $f:X\to\mathbb{R}$ is inf-compact if for every $\lambda\in\mathbb{R}$,
the sublevel set ${\cal L}_{f}\left(\lambda\right)\triangleq\left\{ x\in X:f\left(x\right)\le\lambda\right\} $
is compact.} on $K\left(x\right)$. If the set $K^{\circ}\left(x\right)\triangleq\left\{ u\in\R^{m}:{\cal C}_{g}\left(x,u\right)<0,\forall g\in\left[h\right]\right\} $
is nonempty for every $x\in{\cal O}$, then $\kappa^{*}\left(x\right)\triangleq\underset{u\in K}{\text{arg min\ }}Q\left(x,u\right)$
is continuous.
\end{lem}
The continuity of $\kappa^{*}$ established in Lemma \ref{lem:continuity of k*}
allows Lemma \ref{lem:boundedness of f dot} and Theorem \ref{thm: theta tilde convergence}
to be used in the proof of the following theorem, which presents the
main result of this paper. The implication of the theorem is that
the developed DNN-based optimization problem in (\ref{eq:kappa star})
ensures system trajectories beginning in the user-selected safe set
${\cal S}$ will remain in ${\cal S}$ for all time.
\begin{thm}
\label{thm:main result}Suppose $B:\r^{n}\times\R^{p}\rightarrow\R$
is an aDCBF for the system defined by (\ref{eq:system}) and (\ref{eq:kappa star})
defining a safe set $\S\subseteq\Omega$. Let $\hat{x}$, $\hat{f}$,
and $\hat{\theta}$ update according to (\ref{eq:x hat dot}), (\ref{eq:fhat dot}),
and (\ref{eq:theta hat dot}), respectively. Let the system be initialized
such that $\hat{x}\left(t_{0}\right)=x\left(t_{0}\right)$, $\left\Vert \hat{f}\left(t_{0}\right)\right\Vert \le\overline{f}$,
$z\left(t_{0}\right)\in{\cal Q}$, and $\hat{\theta}\left(t_{0}\right)\in{\cal B}$.
If Assumptions \ref{asm:bounded ideal weights} and \ref{asm:psi}
hold, $Q$ is selected to be continuous, and $u\mapsto Q\left(x,u\right)$
is strictly convex and inf-compact on $K_{d}\left(x\right)$, then
the safe set $\S$ is forward invariant for the closed-loop dynamics
defined by (\ref{eq:system}), the controller $\kappa^{*}$ in (\ref{eq:kappa star}),
and the adaptive weight update law in (\ref{eq:theta hat dot}), provided
$\lambda_{3}>0$.
\end{thm}
\begin{IEEEproof}
Let $t\mapsto x\left(t\right)$ be a solution to the closed-loop dynamics
$f_{cl}$ defined by (\ref{eq:system}) and (\ref{eq:kappa star}).
Because $B$ is an aDCBF in the sense of Definition \ref{def:aDNNCBF},
the set $K_{d}\left(x\right)$ is nonempty on ${\cal O}\supset{\cal N}\left(\S\right)$.
Thus, the optimization-based control law in (\ref{eq:kappa star})
yields a controller $\kappa^{*}\in K_{d}$. Because $Q$ is continuous
and $u\mapsto Q\left(x,u\right)$ is strictly convex and inf-compact
on $K_{d}\left(x\right)$ by assumption, Assumption \ref{asm:psi}
holds, ${\cal C}_{F}$ is continuous on ${\cal O}\times\mathbb{R}^{m}$,
and $u\mapsto{\cal C}_{F}\left(x,u\right)$ is convex on $K_{d}$.
Therefore, the conditions of Lemma \ref{lem:continuity of k*} hold,
thus implying $\kappa^{*}$ is single-valued and continuous on ${\cal O}$.
By Lemma \ref{lem:boundedness of f dot}, there exists a constant
$\bar{\dot{f}}$ such that $\left\Vert \dot{f}\right\Vert \le\bar{\dot{f}}$
for all $x\in{\cal S}$, so Theorem \ref{thm: theta tilde convergence}
can be used to show that the weight update law in (\ref{eq:theta hat dot})
ensures $\left\Vert \tilde{\theta}\right\Vert $ is uniformly ultimately
bounded by (\ref{eq:theta bound}) for all $t\in\mathcal{I}$. Since
$\left\Vert \hat{f}\left(t_{0}\right)\right\Vert \le\overline{f}$,
$z\left(t_{0}\right)\in{\cal D}$, $\hat{\theta}\left(t_{0}\right)\in{\cal B}$,
and $f$ is continuously differentiable, it follows that $\left\Vert \tilde{\theta}\left(t_{0}\right)\right\Vert \le\Xi$,
$\left\Vert \tilde{f}\left(t_{0}\right)\right\Vert \le2\overline{f}$,
and $\left\Vert \Delta\right\Vert \le c_{1}$. Therefore, for every
$u\in\Psi$ and $t\in{\cal I}$, $\nabla B^{\top}\left(x\right)\dot{x}\le\left\Vert \nabla B^{\top}\left(x\left(t\right)\right)\Phi^{\prime}\right\Vert \left(\chi_{\theta}+c_{1}\right)+\nabla B^{\top}\left(x\left(t\right)\right)\left(\Phi\left(x\left(t\right),\hat{\theta}\right)+g\left(x\left(t\right)\right)u\right)$,
implying $\kappa^{*}\left(x\left(t\right)\right)\in K_{d}\left(x\left(t\right)\right)\subset K_{c}\left(x\left(t\right)\right)$
for all $t\in\mathcal{I}$. Since $K_{d}$ is nonempty for all $x\in{\cal O}$,
$K_{c}$ is nonempty on ${\cal O}$ and $B$ is a CBF defining the
set ${\cal S}$ in the sense of Definition \ref{def:CBF}. By Theorem
1 of \cite{Isaly.Mamaghani.ea2024}, ${\cal S}$ is forward pre-invariant
for the closed-loop dynamics defined by (\ref{eq:system}) and (\ref{eq:kappa star}).
Maximal solutions to $f_{cl}$ are either complete or escape in finite-time
by flowing \cite[Proposition 3]{Maghenem.Sanfelice2021}. The safe
set ${\cal S}$ is compact by definition, eliminating the possibility
of finite-time escape from the safe set \cite[Theorem 10.1.4]{Aubin2008},
which implies all maximal solutions to the closed-loop system are
complete, i.e., the maximal $\mathcal{I}=\left[t_{0},\infty\right)$.
Thus, it follows that the safe set ${\cal S}$ is forward invariant.
\end{IEEEproof}

\section{Safety Under Intermittent State Feedback\label{sec:Safety-Under-Intermittent}}

During intermittent loss of feedback, the state measurements are not
available, so it is impossible to use any feedback mechanisms; however,
because the developed DNN yields parameter estimation error guarantees,
it can be used to make state predictions at times when feedback is
lost. Let $k\in\mathbb{Z}_{\geq0}$ denote the time index such that
feedback is unavailable in the time interval $\left[t_{2k+1},t_{2k+2}\right)$.
When feedback is not available, an open-loop estimation of the current
state $\hat{X}\in\r^{n}$ can be updated according to 
\begin{align}
\dot{\hat{X}} & =\Phi\left(\hat{X},\hat{\theta}\left(t_{2k+1}\right)\right)+g\left(\hat{X}\right)u,\label{eq:X hat dot}
\end{align}
where the initial condition for the state estimate is $\hat{X}\left(t_{2k+1}\right)=x\left(t_{2k+1}\right)$.
When feedback becomes available, $\hat{X}\left(t\right)$ is reset
as $\hat{X}\left(t\right)=x\left(t\right)$ for all $\left(t,k\right)\in\left[t_{2k},t_{2k+1}\right)\times\mathbb{Z}_{\geq0}$.
In disturbance observer-based methods such as \cite{Isaly.Patil.ea2024},
the CBF-based constraint is reliant on the observer estimate of the
dynamics and therefore on state measurements, so safety cannot be
ensured when feedback is lost. In the developed approach, the constraint
in (\ref{eq:kappa star}) can be modified to ensure safety until feedback
is restored.

\subsection{Modified CBF Constraint Development}

The open-loop estimator error $\tilde{X}\in\R^{n}$ is defined as
$\tilde{X}\triangleq x-\hat{X}$; thus, $\tilde{X}\left(t_{2k+1}\right)=0$.
Exclusively for the purpose of this section, it is assumed that there
exists a known constant $\bar{u}\in\mathbb{R}^{n}$ such that $\left\Vert u\right\Vert \le\bar{u}$,
for all $u\in\Psi$, the drift $f$ is globally bounded and Lipschitz,
and $g$ is globally Lipschitz. Such an assumption on the boundedness
of $f$ is mild since finite-time escape is not inherent to the uncontrolled
dynamics for most physical systems of practical interest. Based the
assumptions on the boundedness of $u$ and $f$ and the continuous
differentiability of $\Phi$, there exists a Lipschitz constant $L_{U}\in\R_{>0}$
such that 
\begin{align}
\left\Vert \Phi\left(x,\theta^{*}\right)-\Phi\left(\hat{X},\theta^{*}\right)+\left(g\left(x\right)-g\left(\hat{X}\right)\right)u\right\Vert  & \leq L_{U}\left\Vert \tilde{X}\right\Vert ,\label{eq:L_U bound}
\end{align}
and, additionally, there exists a constant $\Delta_{U}\in\R_{>0}$
such that 
\begin{align}
\left\Vert \Phi\left(\hat{X},\theta^{*}\right)-\Phi\left(\hat{X},\hat{\theta}\left(t_{2k+1}\right)\right)+\varepsilon\left(x\right)\right\Vert  & \le\Delta_{U}.\label{eq:Delta_U}
\end{align}
Let the Lyapunov function candidate during loss of feedback $V_{U}:\R^{2n+p}\rightarrow\R$
be defined as $V_{U}\triangleq\frac{1}{2}\tilde{X}^{\top}\tilde{X}$.
Taking the time-derivative of $V_{U}$, adding and subtracting $\Phi\left(\hat{X},\theta^{*}\right)$,
substituting in (\ref{eq:L_U bound}) and (\ref{eq:Delta_U}), and
using Young's Inequality, it can be shown that $\dot{V}_{U}\le\lambda_{U}V_{U}+\frac{\Delta_{U}}{2}$
when feedback is unavailable, where $\lambda_{U}\triangleq2L_{U}+\Delta_{U}$.
Solving for $V_{U}$ yields $V_{U}\left(t\right)\le\left(V\left(t_{2k+1}\right)+\delta_{U}\right)e^{\lambda_{U}\left(t-t_{2k+1}\right)}-\delta_{U}$
for all $\left(t,k\right)\in\left[t_{2k+1},t_{2k+2}\right)\times\mathbb{Z}_{\ge0}$,
where $\delta_{U}\triangleq\frac{2\Delta_{U}}{2L_{U}+\Delta_{U}}$.
Therefore, the open-loop estimation error dynamics can be shown to
satisfy the form $\left\Vert \dot{\tilde{X}}\right\Vert \le L_{U}\left\Vert \tilde{X}\right\Vert +\Delta_{U}$,
which is exponentially unstable (cf., \cite{Chen.Bell.ea2019,Parikh.Cheng.ea2017})
such that 
\begin{align}
\left\Vert \tilde{X}\left(t\right)\right\Vert  & \le\overline{\tilde{X}}\left(t\right)\triangleq\sqrt{\delta_{U}\left(e^{\lambda_{U}\left(t-t_{2k+1}\right)}-1\right)}\label{eq:X tilde}
\end{align}
for all $\left(t,k\right)\in\left[t_{2k+1},t_{2k+2}\right)\times\mathbb{Z}_{\ge0}$.
The system dynamics can be bounded as $\left\Vert \dot{x}\right\Vert \le\left\Vert \dot{\tilde{X}}\right\Vert +\left\Vert \dot{\hat{X}}\right\Vert \le L_{U}\overline{\tilde{X}}+\Delta_{U}+\left\Vert \Phi\left(\hat{X},\hat{\theta}\left(t_{2k+1}\right)\right)+g\left(x\right)u\right\Vert $,
and if $\nabla B$ is locally Lipschitz, then $\left\Vert \nabla B^{\top}\left(x\right)-\nabla B^{\top}\left(\hat{X}\right)\right\Vert \le\rho\overline{\tilde{X}}\left(t\right)$,
where $\rho\in\R_{>0}$ is a positive constant. Thus, the controller
in (\ref{eq:kappa star}) in absences of state feedback becomes
\begin{align}
\kappa^{*}\left(\hat{X}\right) & \triangleq\underset{u\in\Psi}{\arg\min}\ Q\left(\hat{X},u\right),\nonumber \\
 & \ \text{s.t.}\ \mathcal{C}_{U}\left(\hat{X},u\right)\le0,\label{eq:no feedback controller}
\end{align}
where{\small{}
\begin{align}
\mathcal{C}_{U}\left(\hat{X},u\right) & \triangleq\overline{\tilde{X}}\left(t\right)\rho\biggl(L_{U}\overline{\tilde{X}}\left(t\right)+\Delta_{U}+\left\Vert \Phi\left(\hat{X},\hat{\theta}\left(t_{2k+1}\right)\right)\right.\nonumber \\
+ & \left.\left.g\left(\hat{X}\right)u\right\Vert \right)+\left\Vert \nabla B^{\top}\left(\hat{X}\right)\right\Vert \biggl(L_{U}\overline{\tilde{X}}\left(t\right)+\Delta_{U}\biggr)\nonumber \\
+ & \nabla B^{\top}\left(\hat{X}\right)\left(\Phi\left(\hat{X},\hat{\theta}\left(t_{2k+1}\right)\right)+g\left(\hat{X}\right)u\right)+\gamma\left(\hat{X}\right).\label{eq:no feedback CBF constraint}
\end{align}
}The development in Sections \ref{sec:Problem-Formulation} and \ref{sec:Stability-Analysis}
rely on the continuity of the control input $u$. When feedback is
lost, the controller switches, introducing discontinuities. The constraint
in (\ref{eq:no feedback CBF constraint}) developed for intermittent
loss of feedback is empirically tested in one of the following simulations.
To account for the effects of switching, the following subsection
introduces a switched systems analysis for CBFs with a dwell-time
condition.

\subsection{Maximum Loss of Feedback Dwell-Time Condition}

As the time without feedback increases, the worst-case bound on $\tilde{X}$
grows exponentially, causing the constraint in (\ref{eq:no feedback CBF constraint})
to shrink the system's operating region. To ensure the feasibility
of the controller with the modified constraint in (\ref{eq:no feedback CBF constraint}),
a maximum loss of feedback dwell-time condition $\Delta t_{k}\in\mathbb{R}_{>0}$
can be developed using (\ref{eq:X tilde}). In this subsection, it
is additionally assumed that $B$ is globally Lipschitz\footnote{While a global Lipschitzness requirement may appear restrictive, many
safe sets of practical interest such as polytopes (i.e., sets described
by the intersection of hyperplanes) can be described using globally
Lipschitz CBF candidates.}, implying that $\left\Vert \nabla B\left(x\right)\right\Vert \le\bar{B}$,
where $\bar{B}\in\mathbb{R}_{>0}$ is a known positive constant. To
ensure the existence of a safety-ensuring control input, the time
without feedback must be such that the inequality 
\begin{align}
\mathcal{C}_{U}\left(\hat{X},u\right)-\mathcal{C}^{*}\left(x,u\right) & \le\bar{\mathcal{C}}\label{eq:dwell time constraint}
\end{align}
is satisfied, where $\mathcal{C}^{*}\left(x,u\right)\triangleq\nabla B^{\top}\left(x\right)\dot{x}$
and $\bar{\mathcal{C}}\in\mathbb{R}^{d}$ is the user-selected maximum
offset between the boundary of the safe set and the boundary of the
operating region enforced by $\mathcal{C}_{U}$. Recall that, for
the purpose of this section, the control input is assumed to be bounded
such that $\left\Vert u\right\Vert \le\bar{u}$, for all $u\in\Psi$.
Substituting (\ref{eq:no feedback CBF constraint}) into (\ref{eq:dwell time constraint})
and solving for $\Delta t_{k}\triangleq t_{2k+1}-t_{2k+2}$ yields
a maximum loss of feedback dwell-time condition of
\begin{align}
\Delta t_{k} & \le\frac{1}{\lambda_{U}}\text{ln}\left(\frac{1}{\delta_{U}}\left(\left(\frac{\bar{\mathcal{C}}-6\bar{B}\Delta_{U}-\mathcal{K}_{U}}{6L_{U}\bar{B}}\right)^{2}+1\right)\right),\label{eq:dwell-time condition}
\end{align}
where $\mathcal{K}_{U}\triangleq4\bar{B}\left\Vert \Phi\left(\hat{X},\hat{\theta}\left(t_{k}\right)\right)\right\Vert +4\bar{B}\left\Vert g\left(\hat{X}\right)\right\Vert \bar{u}$.

\section{Simulation Studies\label{sec:Simulation-Studies}}

Two simulations are provided to demonstrate the effectiveness of the
developed aDCBFs. The optimization problem in (\ref{eq:kappa star})
is used to define the control law with a cost function of $Q\left(x,u\right)=\left\Vert u-u_{nom}\left(x\right)\right\Vert ^{2}$
, where $u_{nom}\in\Psi$ is the nominal control input that tracks
the desired trajectory. 

\subsection{Adaptive Cruise Control\label{subsec:Adaptive-Cruise-Control}}

\begin{figure}
\centering{}\includegraphics[width=1\columnwidth]{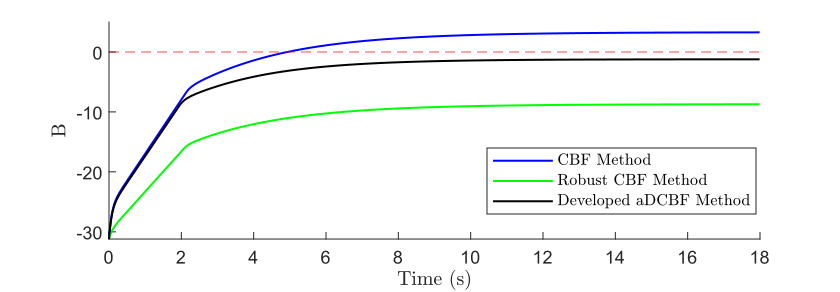}\caption{\label{fig: safe set}The value of the barrier functions over time
for the ACC problem. A negative value of $B$ indicates the follower
vehicle remains in the safe set.}
\end{figure}
In this section, the developed technique is applied to an adaptive
cruise control (ACC) problem \cite{Ames.Xu.ea2016}. Suppose there
are two vehicles traveling along a straight line. The lead vehicle
travels forward with a velocity $v_{\text{lead}}\in\mathbb{R}$ of
$v_{\text{lead}}=10$ m/s, while the follower vehicle trails behind
the lead vehicle. The follower vehicle has dynamics
\begin{align*}
\left[\begin{array}{c}
\dot{x}\\
\dot{v}
\end{array}\right] & =\left[\begin{array}{c}
v\\
-\frac{1}{m}F_{r}\left(v\right)+\delta\left(v\right)
\end{array}\right]+\left[\begin{array}{c}
0\\
\frac{1}{m}
\end{array}\right]u,
\end{align*}
where $x\in\mathbb{R}$ is the position of the vehicle in meters,
$v\in\mathbb{R}$ is the velocity of the vehicle in meters per second,
$m\in\mathbb{R}$ is the mass in kg, the nonlinear function $F_{r}:\mathbb{R}\rightarrow\mathbb{R}$
represents the vehicle's rolling resistance, the function $\delta:\mathbb{R}\rightarrow\mathbb{R}$
represents an unknown disturbance, and $u\in\mathbb{R}$ is the control
input. As described in \cite{Ames.Xu.ea2016}, the rolling resistance
is modeled as $F_{r}\left(v\right)=f_{0}+f_{1}v+f_{2}v^{2}$, where
$f_{0}=0.1$ N, $f_{1}=5\ \text{\ensuremath{\frac{\text{N\ensuremath{\cdot}s}}{\text{m}}}}$,
and $f_{2}=0.25\ \frac{\text{N\ensuremath{\cdot s^{2}}}}{\text{m}}$.
The added disturbance $\delta\left(v\right)=30\sin\left(0.1v\right)$
represents unmodeled forces on the vehicle and the mass of the vehicle
is $m=100$ kg. In the aDCBF method, the rolling resistance function
is considered to be unknown and $f\left(x\right)=-\frac{1}{m}F_{r}\left(v\right)+\delta\left(v\right)$
is the nonlinear function that the DNN learns. The desired velocity
of the follower vehicle $v_{d}\in\mathbb{R}$ is set to a constant
$v_{d}\triangleq20$ m/s. The distance between the lead and follower
vehicles $D\in\mathbb{R}$ is defined as $D\triangleq x_{\text{lead}}-x$,
where $x_{\text{lead}}\in\mathbb{R}$ is the position of the lead
vehicle. The vehicles are initialized such that $v\left(t_{0}\right)=16$
m/s, $v_{\text{lead}}\left(t_{0}\right)=10$ m/s, and $D\left(t_{0}\right)=60$
m. Because $v_{\text{lead}}<v_{d}$, the nominal velocity tracking
controller defined as $u_{nom}\triangleq-\Phi\left(v,\hat{\theta}\right)-mk_{1}\left(v-v_{d}\right)$,
where $k_{1}=10$ is a user-selected control gain, would eventually
cause the follower vehicle to collide with the leader. The developed
aDCBF approach is used to enforce a safe following distance. The safe
set is defined as $\mathcal{S}\triangleq\left\{ v\in\R:B\left(v\right)=-D+1.8v\le0\right\} $,
where 1.8 s represents the desired time headway as in \cite{Ames.Xu.ea2016}.
A deep ResNet is used with 2 hidden layers, a shortcut connection
between each layer, and 6 neurons in each layer, for a total of 122
individual layer weights. The weights are initialized from the normal
distribution $N\left(0,3\right)$ and the DNN gains are selected as
$\Gamma\left(t_{0}\right)=5I_{122}$, $k_{\theta}=0.001$, $\beta_{0}=2$,
and $\kappa_{0}=3$. The state-derivative estimator is used to produce
the secondary estimate of $f$ uses gains of $k_{x}=5$ and $k_{f}=10$.
Figure \ref{fig: safe set} demonstrates how the controller in (\ref{eq:kappa star})
with $\gamma\left(v\right)=10B\left(v\right)$ is able to constrain
the follower vehicle to a safe distance behind the lead vehicle. 

Results for two comparison simulations are also provided in Figure
\ref{fig: safe set}. Using a standard CBF approach \cite{Ames.Xu.ea2016},
the nominal controller is given access to $F_{r}$ and $m$ but does
not have information about the disturbance term, meaning $u_{nom}\triangleq F_{r}\left(v\right)-mk_{1}\left(v-v_{d}\right)$
and the CBF constraint is defined as $\mathcal{C}_{F}\left(v,u\right)\triangleq\dot{D}+1.8\left(-\frac{1}{m}F_{r}\left(v\right)+\frac{1}{m}u\right)+10B\left(v\right)$.
The unmodeled uncertainty $\delta$ pushes the state trajectory out
of the safe set ($B$ reaches a positive steady-state value of 3.15,
thus violating the safe following distance requirements). If it is
known that the model of $F_{r}$ is imperfect, a robust CBF approach
can be used, with $u_{nom}\triangleq F_{r}\left(v\right)-m\bar{\delta}-mk\left(v-v_{d}\right)$
and $\mathcal{C}\left(v,u\right)\triangleq\dot{D}+1.8\left(-\frac{1}{m}F_{r}\left(v\right)+\bar{\delta}+\frac{1}{m}u\right)+10B\left(v\right)$,
where $\bar{\delta}\in\mathbb{R}_{>0}$ is a known constant such that
$\left\Vert \delta\right\Vert \le\bar{\delta}$. Although the robust
approach is able to keep the trajectory inside the safe set, the use
of a worst-case bound on $\delta$ results in an overly conservative
set of admissible controllers, restricting the state trajectory to
a subset of the safe set. Using the robust CBF approach, $B$ reaches
a steady-state value of $-8.81$. Adaptive CBF methods such as those
in \cite{Taylor.Ames2020} and \cite{Lopez.Slotine.ea2020} cannot
be directly applied to this problem because of the nonlinearly parameterized
uncertainty in $\delta$. Using the developed method, $B$ reaches
a stead state value of $-1.27$. The developed aDCBF method ensures
safety while reducing undesirable conservative behavior by $85.6$\%,
compared to robust CBFs.

\subsection{Non-Polynomial Dynamics\label{sec:Example}}

\begin{figure}
\centering{}\includegraphics[width=1\columnwidth]{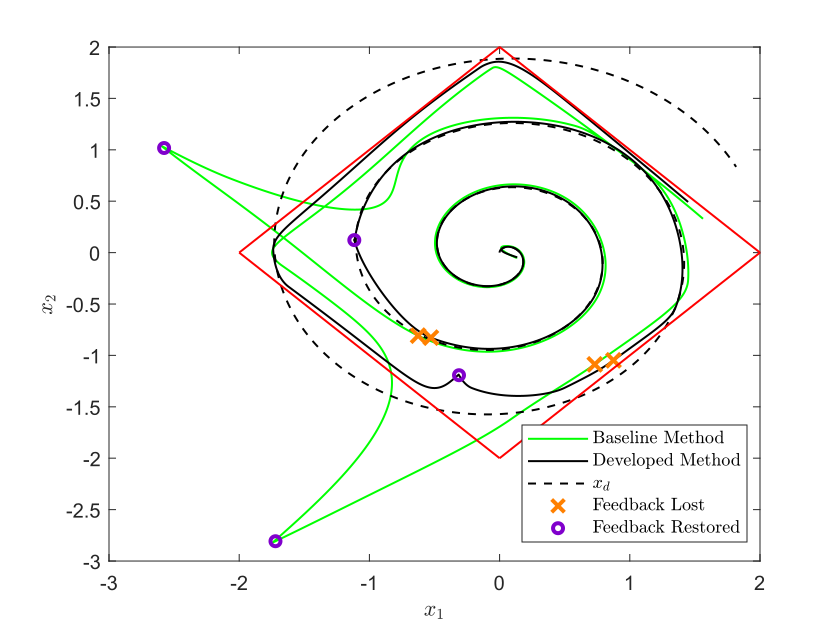}\caption{\label{fig:The-state-trajectory}The state trajectory of the closed-loop
system in Section \ref{sec:Example} using the developed aDCBF approach
(black line) compared to the same control scheme without the ResNet
approximation of the dynamics (green line).}
\end{figure}

Consider the nonlinear dynamical system in (\ref{eq:system}) with
$f\left(x\right)=\left[x_{2}\sin\left(x_{1}\right)\tanh^{2}\left(x_{2}\right),\ x_{1}x_{2}\cos\left(x_{2}\right)\text{sech}\left(x_{2}\right)\right]^{\top}$
and $g\left(x\right)=\left[1,\ 1\right]^{\top}$, where $x=\left[x_{1},x_{2}\right]^{\top}.$
White Gaussian noise was added to the position state measurement with
a signal to noise ratio of 50 dB. The position state is initialized
from the uniform distribution $U\left(-0.2,0.2\right)$ with $\dot{x}\left(t_{0}\right)=\left[0,\ 0\right]^{\top}$,
and the desired trajectory is defined as $x_{d}\left(t\right)=0.1t\left[\sin\left(t\right),\ \cos\left(t\right)\right]^{\top}$.
We define a vector-valued CBF as 
\begin{align*}
B\left(x\right) & \triangleq\left[\begin{array}{c}
x_{1}+x_{2}-2\\
x_{1}-x_{2}-2\\
-x_{1}+x_{2}-2\\
-x_{1}-x_{2}-2
\end{array}\right],
\end{align*}
which defines a diamond safe set $\S=\left\{ x\in\R^{2}:B\left(x\right)\le0\right\} $
with height and width of 4. The deep ResNet has 3 hidden layers, a
shortcut connection across each hidden layer, and with 5 neurons in
each layer, thus involving 174 total weights. The ResNet\textcolor{blue}{{}
}weights are initialized from the normal distribution $N\left(0,0.5\right)$
and the DNN gains are selected as $\Gamma\left(t_{0}\right)=5I_{174}$,
$\alpha=50,$ $k_{\theta}=0.001$, $\beta_{0}=2$, and $\kappa_{0}=10$.
The state-derivative estimator is used to produce the secondary estimate
of $f$ using gains of $k_{x}=10$ and $k_{f}=5$. The nominal controller
is defined as $u_{nom}=\dot{x}_{d}-\Phi\left(x,\hat{\theta}\right)-k_{e}\left(x-x_{d}\right)$,
where $k_{e}=10$. The function $\gamma$ is selected as $\gamma\left(x\right)=10B\left(x\right)$.
Figure \ref{fig:The-state-trajectory} shows the safe set and the
desired and actual state trajectories. The simulation was run for
20 seconds with a step size of 0.005 seconds.\textcolor{blue}{{} }To
simulate performance under intermittent state feedback, the state
measurement is made unavailable for 1 second intervals beginning at
10 seconds and 15 seconds, denoted by the orange $\boldsymbol{\times}$
markers in Figure \ref{fig:The-state-trajectory}. During loss of
feedback, the procedure in Section \ref{sec:Safety-Under-Intermittent}
is followed, where the identified DNN is used to make a prediction
of the state and the modified robust CBF constraint in (\ref{eq:no feedback CBF constraint})
is used to ensure safety. The method developed in Section \ref{sec:Safety-Under-Intermittent}
prevents the state from escaping the safe set until the feedback is
restored at 11 seconds and 16 seconds, respectively, indicated by
the purple $\boldsymbol{\circ}$ markers.

\begin{figure}
\centering{}\includegraphics[width=1\columnwidth]{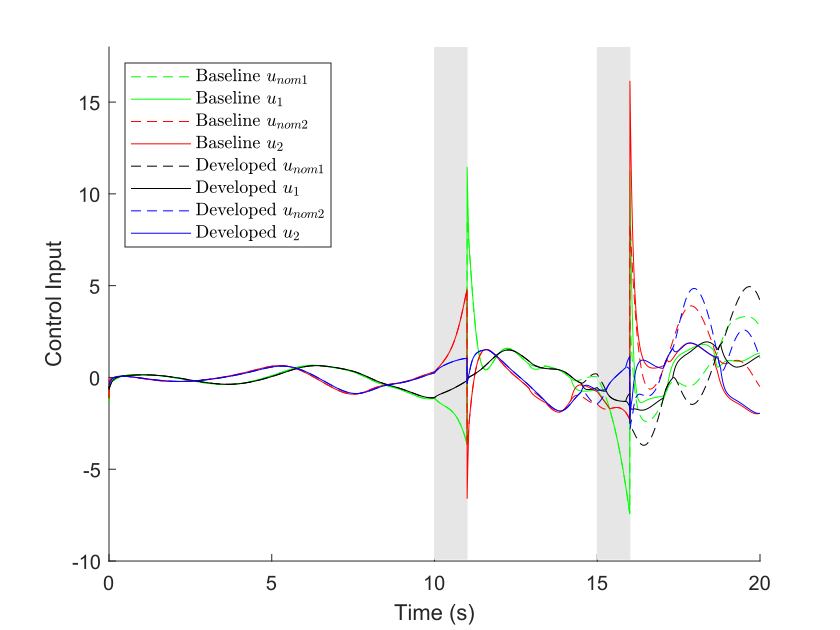}\caption{\label{fig:control input}Comparison plots of the control inputs for
the developed and baseline methods. The gray regions represent the
time periods where state feedback is made unavailable.}
\end{figure}

For the baseline method in \cite{Isaly.Patil.ea2024}, the nominal
controller is $u_{nom}\triangleq\dot{x}_{d}-\hat{d}-k_{e}\left(x-x_{d}\right)$,
where $\hat{d}\in\mathbb{R}^{2}$ represents the RISE-based disturbance
observer estimate of $f$ defined in \cite[Eq. 6]{Isaly.Patil.ea2024},
and the CBF constraint in the optimization-based controller is $\mathcal{C}_{F}\left(x,u\right)\triangleq\left\Vert \nabla B^{\top}\left(x\right)\right\Vert \overline{f}+\nabla B^{\top}\left(x\right)g\left(x\right)u+\gamma\left(x\right)$.
Though trajectory tracking performance is comparable to the developed
approach at times when feedback is available, the state-derivative
observer alone only provides an instantaneous estimate of the dynamics
and fails to ensure safety in both instances of feedback loss, unlike
the developed method. The nominal and implemented control inputs for
the baseline and developed methods are shown in Figure \ref{fig:control input}.
It can be seen that the developed controller in (\ref{eq:kappa star})
with the selected cost function acts as a safety filter, modifying
the nominal control input when the states get close to the boundary
of the safe set and otherwise returning the same values as the nominal
control input. Additionally, it can be seen that when feedback is
made unavailable, the baseline controller reacts more aggressively
causing spikes in the control input. The aggressive control action
and violation of safety constraints are likely because the observer-based
estimate of the dynamics is inaccurate under the loss of feedback.
This inaccuracy is because, unlike the DNN model, the observer does
not have the capacity to generalize beyond the explored trajectory
points. However, because the developed method uses a DNN model that
learns the dynamics online, the learned DNN is able to generalize
on unseen trajectory points encountered during the loss of feedback.
As a result, the optimization-based controllers in (\ref{eq:kappa star})
and (\ref{eq:no feedback controller}) ensure the safety constraints
are satisfied while being less aggressive than the baseline controller.

\begin{figure}
\centering{}\includegraphics[width=1\columnwidth]{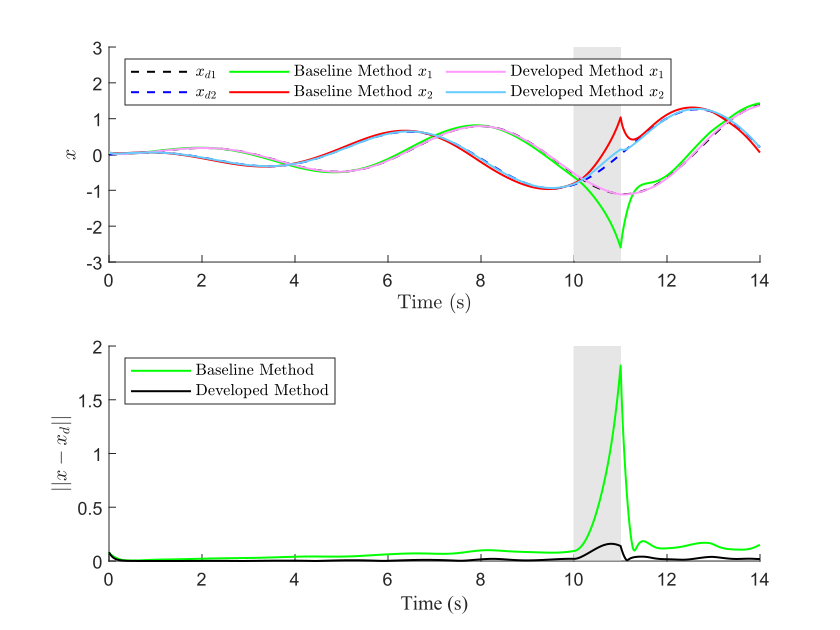}\caption{\label{fig:error trajectory}The top plot shows the desired versus
actual value of each position state for the baseline and developed
methods over the first 14 seconds of the simulation. The bottom plot
shows the position tracking error for the two methods over the first
14 seconds of the simulation. The gray region represents the time
period where state feedback is made unavailable.}
\end{figure}

Figure \ref{fig:error trajectory} shows the trajectory tracking performance
of the two methods. The position tracking error for the developed
method spikes at the start of the simulation, which can be accredited
to the random initialization of the DNN weights. The developed weight
adaptation law in (\ref{eq:weight update law}) helps to enable the
position error to settle in less than $0.5$ seconds. When feedback
is lost at $10$ seconds, the position error norm grows to a maximum
value of $0.56$ with the developed method compared to a value of
$1.81$ with the baseline method, thus achieving a $69.1\%$ tracking
performance improvement. The root mean square position error norm
between $0$ and $14$ seconds is \textbf{$0.035$ }with the developed
method compared to $0.256$ with the baseline method, thus achieving
a $86.3\%$ tracking performance improvement. The position tracking
error from 14 seconds to 20 seconds is omitted from the plot because
the desired trajectory in this timespan is outside of the safe set,
rendering position error uninformative.

\begin{table*}
\centering{}\caption{\label{tab:results-1}Performance Results Over 50 Iterations}
\begin{tabular}{|c|c|c|c|>{\centering}m{0.25\columnwidth}|}
\hline 
Control Method &
Trajectory &
Max. $B$ &
Avg. Max. $B$ &
Avg. Time \\
Outside ${\cal S}$ (s)\tabularnewline
\hline 
\hline 
\multirow{4}{*}{Developed} &
Spiral 1 &
-0.18 &
-0.38 &
0\tabularnewline
\cline{2-5} \cline{3-5} \cline{4-5} \cline{5-5} 
 & Spiral 2 &
-0.13 &
-0.24 &
0\tabularnewline
\cline{2-5} \cline{3-5} \cline{4-5} \cline{5-5} 
 & Figure Eight &
-0.05 &
-0.51 &
$0$\tabularnewline
\cline{2-5} \cline{3-5} \cline{4-5} \cline{5-5} 
 & Average &
-0.12 &
-0.38 &
0\tabularnewline
\hline 
\multirow{4}{*}{Baseline} &
Spiral 1 &
52.83 &
5.84 &
0.40\tabularnewline
\cline{2-5} \cline{3-5} \cline{4-5} \cline{5-5} 
 & Spiral 2 &
18.36 &
3.40 &
0.11\tabularnewline
\cline{2-5} \cline{3-5} \cline{4-5} \cline{5-5} 
 & Figure Eight &
91.14 &
8.55 &
$0.57$\tabularnewline
\cline{2-5} \cline{3-5} \cline{4-5} \cline{5-5} 
 & Average &
54.11 &
5.93 &
0.36\tabularnewline
\hline 
\end{tabular}
\end{table*}
 To demonstrate consistent safety performance across various reference
trajectories and initial conditions, the simulations were repeated
for 50 iterations for three different desired trajectories. The three
desired trajectories tested were the previously used spiral trajectory
$x_{d}\left(t\right)=0.1t\left[\sin\left(t\right),\ \cos\left(t\right)\right]^{\top}$(henceforth
referred to as ``Spiral 1''), a second spiral trajectory defined
as $x_{d}\left(t\right)=0.075t\left[\sin\left(t\right),\ \cos\left(t\right)\right]^{\top}$(henceforth
referred to as ``Spiral 2''), and a figure-eight trajectory defined
as $x_{d}\left(t\right)=\left[2\sin\left(t\right),\ 2\sin\left(t\right)\cos\left(t\right)\right]^{\top}$.
For each of the iterations for each of the three trajectories, the
initial states were again randomly selected from the uniform distribution
$U\left(-0.2,0.2\right)$ with $\dot{x}\left(t_{0}\right)=\left[0,\ 0\right]^{\top}$,
and the DNN weights were randomly initialized from the normal distribution
$N\left(0,0.5\right)$. The gains were kept the same as previously
reported. Loss of feedback was simulated for two one-second periods,
with the first instance of loss of feedback beginning at a time selected
from the uniform distribution $U\left(0,9\right)$ and the second
instance of loss of feedback beginning at a time selected from the
uniform distribution $U\left(10,19\right)$. The performance results
for each desired trajectory are shown in Table \ref{tab:results-1}.
As can be seen in Table \ref{tab:results-1}, the maximum value of
$B$ for each trajectory across all iterations remains negative using
the developed method. The negative value of $B$ indicates that the
developed aDCBF method successfully maintained the state inside the
safe set at all times over each trajectory and each iteration, whereas,
in the baseline method, the average time spent outside of the safe
set over all three trajectories is $0.36$ seconds, which is $18$\%
of the time feedback was made unavailable. The developed aDCBF method
yielded similar results across all iterations regardless of the desired
trajectory used or differences in the initial DNN weights.

\section{Conclusion\label{sec:Conclusion}}

This paper provides a method to ensure safety while learning the system's
uncertain dynamics in real-time. The developed method is the first
result that combines CBFs with an adaptive DNN that updates in real-time,
eliminating the need for pre-training. Unlike previous Lb-DNN literature
where the DNN adaptation laws are based on a tracking error, the developed
DNN adaptation law is based on an identification error. The developed
DNN adaptation law yields function approximation error convergence,
which iss then used in an optimization-based control law that enforced
forward invariance of the safe set. A combined Lyapunov-based analysis
is proven to achieve guarantees on the DNN function approximation
under the PE condition. Furthermore, to ensure safety under intermittent
loss of feedback, the developed aDCBF is extended by leveraging the
learned DNN's capacity to extrapolate beyond the explored trajectory
points. A switched systems analysis for CBFs is provided with a maximum
dwell-time condition during which the feedback can be unavailable.
Simulation results are provided for ACC and intermittent feedback
system examples with comparisons to robust CBFs and observer-based
CBFs as baselines. The developed aDCBF method ensured safety while
reducing undesirable conservative behavior by $85.6$\%, compared
to a robust CBF approach. Future work may involve extension to high-order
aDCBFs that ensure safety for systems of higher relative degree.

\bibliographystyle{IEEEtran}
\bibliography{5C__Users_hsweatland_Documents_Papers_ncrbibtex_bibtex_bib_ncrbibs_encr,6C__Users_hsweatland_Documents_Papers_ncrbibtex_bibtex_bib_ncrbibs_master,7C__Users_hsweatland_Documents_Papers_ncrbibtex_bibtex_bib_ncrbibs_ncr}

\begin{IEEEbiography}[{\includegraphics[scale=0.21]{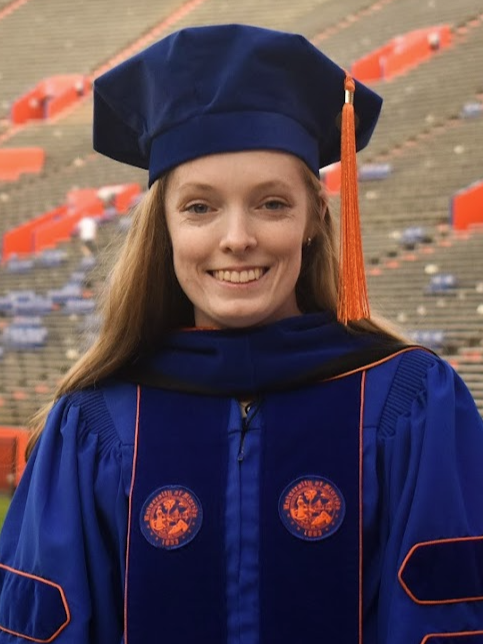}}]{Hannah M. Sweatland}
 received B.S, M.S., and Ph.D. degrees in mechanical engineering
from the University of Florida, Gainesville, FL, USA in 2020, 2022,
and 2024, respectively. Her research interests include safety of nonlinear
systems, adaptive control, and passivity-based control.
\end{IEEEbiography}

\begin{IEEEbiography}[{\includegraphics[scale=0.55]{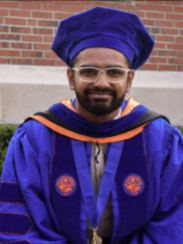}}]{Omkar Sudhir Patil}
 received his Bachelor of Technology (B.Tech.) degree in production
and industrial engineering from Indian Institute of Technology (IIT)
Delhi in 2018, where he was honored with the BOSS award for his outstanding
bachelor's thesis project. In 2019, he joined the Nonlinear Controls
and Robotics (NCR) Laboratory at the University of Florida under the
guidance of Dr. Warren Dixon to pursue his doctoral studies. Omkar
received his Master of Science (M.S.) degree in mechanical engineering
in 2022 and Ph.D. in mechanical engineering in 2023 from the University
of Florida. During his PhD studies, he was awarded the Graduate Student
Research Award for outstanding research. In 2023, he started working
as a postdoctoral research associate at NCR Laboratory, University
of Florida. His research focuses on the development and application
of innovative Lyapunov-based nonlinear, robust, and adaptive control
techniques.
\end{IEEEbiography}

\begin{IEEEbiography}[{\includegraphics[scale=0.225]{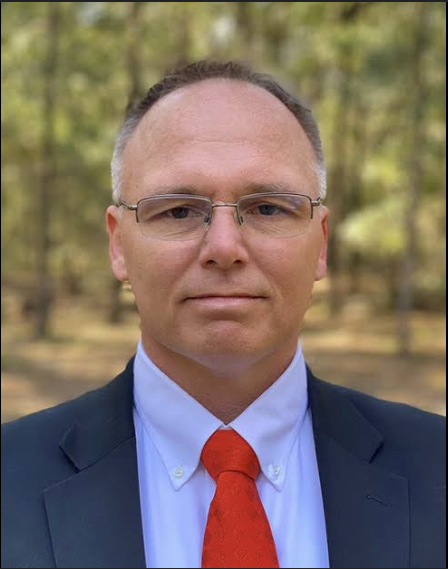}}]{Warren E. Dixon}
 received the Ph.D. degree in electrical engineering from Clemson
University, Clemson, SC, USA, in 2000. He joined the Universidy of
Florida, Gainesville, FL, USA, in 2004 and is now the Distinguished
Professor, Dean's Leadership Professor, and Department Chair with
the Department of Mechanical Aerospace Engineering. He was a Wigner
Fellow and Research Staff Member with the Oak Ridge National Laboratory,
Oak Ridge, TN, USA. His main research interest has been the development
and application of Lyapunov-based control techniques for uncertain
nonlinear systems. Dr. Dixon was the recipient of various early and
mid-career awards and best paper awards. He is an ASME Fellow for
contributions to adaptive control of uncertain nonlinear systems.
\end{IEEEbiography}

\end{document}